\pdfoutput=1
\documentclass[%
reprint,
onecolumn,
preprint,
amsmath,amssymb,
aps,
prfluids
]{revtex4-2}
\usepackage{graphicx}

\usepackage{caption}
\usepackage{subcaption}
\usepackage{dcolumn}
\usepackage{bm}

\usepackage[bookmarks=false]{hyperref}
\usepackage{indentfirst}
\usepackage{amsmath}
\usepackage{amssymb}
\usepackage{soul}
\usepackage{makecell}
\usepackage{tikz}

 \usepackage{setspace}
 \usepackage{csvsimple}
 \usepackage{xcolor}
\makeatletter
\def\mathcolor#1#{\@mathcolor{#1}}
\def\@mathcolor#1#2#3{%
  \protect\leavevmode
  \begingroup
    \color#1{#2}#3%
  \endgroup
}
\makeatother
 
\usepackage[normalem]{ulem}

\begin{document}

\preprint{APS/123-QED}

\title{Drag and torque coefficients of a translating particle with slip at a gas-liquid interface}
\author{Zhi Zhou}
\author{Petia M. Vlahovska}%
\email{petia.vlahovska@northwestern.edu}
\author{Michael J. Miksis}%
 \email{miksis@northwestern.edu}
\affiliation{%
Department of Engineering Sciences and Applied Mathematics, Northwestern University\\
Evanston, IL 60208 
}

\date{\today}

\begin{abstract}
The dynamics of colloid-size particles trapped at a
liquid interface  is an  extensively studied problem owing to its relevance to a wide range of engineering applications. Here we investigate the impact of interfacial deformations on the hydrodynamic force and torque exerted on a  spherical particle with surface slip 
moving along a gas-liquid interface. Following a two-parameter asymptotic modeling approach, we perturb the interface from its planar state and apply the Lorentz reciprocal theorem to the zeroth and first-order approximations to analytically calculate the drag and torque on the particle. 
This allows us to explicitly account for the effect of physical parameters like the three-phase contact angle, the Bond number, and the slip coefficient on the particle motion. In addition, we study the  interactions between two translating and rotating particles at a large separation. The interaction forces and torques exerted by the flow-induced deformations  are calculated via the linear superposition approximation, where the interaction forces are identified as dipolar in terms of the azimuthal angle. 
\end{abstract}

\maketitle


\section{Introduction}
Particles attached to an interface separating two immiscible fluids have been a subject of intense research. In the case of colloid-sized particles, their absorption onto the fluid interface minimizes the interfacial energy, which consequently traps the particles and restricts their motion to only translation and rotation along the interface \cite{binks2006,Leal1980,Maldarelli2022}. 
The motion of a particle translating along  planar and curved interfaces has been examined in several studies \cite{Dani2015,Danov1995, Danov2000,Das2018,dorr2015,dorr2016,fischer2006, Pozrikidis2007, Hu2021}. The drag coefficients of a particle trapped at an interface at varying immersion depths have been evaluated numerically through the use of boundary integral methods \cite{Pozrikidis2007,fischer2006}, finite-difference methods \cite{Danov1995, Danov2000}, and finite element methods \cite{Das2018,Loudet2020}, assuming that the particle does not rotate. Analytical calculations using the integral transform methods were performed by Dani \textit{et al.} \cite{Dani2015}, D\"orr \textit{et al.} \cite{dorr2015} and D\"orr and Hardt \cite{dorr2016} under the assumption of contact angle hysteresis that pins the three-phrase contact line (TCL) to the surface of the translating particle. Loudet \textit{et al.} \cite{Loudet2020} and Zhou \textit{et al.} \cite{zhou2022} have investigated the effect of interfacial deformations on the drag coefficient of a translating particle, assuming a constant contact angle along the TCL. Loudet \textit{et al.}  \cite{Loudet2020} computed the drag coefficient of a two-dimensional (2D) circular cylinder straddling a deformable interface using a phase-field model. Zhou \textit{et al.} \cite{zhou2022} derived analytical expressions for the drag coefficient of one and two spherical particles at a deformable interface using a combined asymptotic and numerical approach. 

As the particle translates along an interface between two immiscible fluids, the asymmetric stress distribution over the particle surface generates a hydrodynamic torque which can cause the particle to rotate.
D\"orr and Hardt \cite{dorr2015} studied the problem where the  particle is rotated by a small angle until the viscous and surface tension torques  balance. Due to the assumption of a pinned TCL, the plane containing the TCL rotates with the particle. The resulting interfacial deformation  is then identified as dipolar \cite{dorr2015}. For a continuously rotating particle, a nonintegrable stress singularity arises at the TCL if no-slip is assumed at the particle surface.  The nonintegrable contact line singularity can be removed by allowing the fluid to slip at the particle surface \cite{Dussan-1974,Haley-1991,Lauga2007}. Adopting the slip boundary condition, O'Neill \textit{et al.} \cite{oneill1986} obtained analytical solutions for the drag and torque coefficients of a rotating sphere at a gas-liquid interface in the limit of a nondeformable flat interface and 90 degree contact angle. In the no-slip limit, the torque caused by particle rotation diverges logarithmically. Das \textit{et al.} \cite{Das2018} carried out numerical calculations of the drag and torque exerted on a rotating particle in the limit of a nondeformable flat interface at different immersion depths using finite element simulations.

Beyond an isolated particle, the interactions mediated by surface deformation, i.e., the capillary interactions, between particles have been studied \cite{Binks2002,Chan1981,Dani2015, Danov2005,Dominguez2008, dorr2015,Ghezzi2001,Hemauer2021,Kralchevsky2000,Nicolson1949,Oettel2005,Oettel2008,Singh2005,Stamou2000,Vassileva2005}. A popular approach is to use the  linear superposition approximation (LSA) introduced by Nicolson \cite{Nicolson1949}.  D\"orr and Hardt \cite{dorr2015} studied the capillary interactions between two translating particles at large separation  and discovered  that the driven particles behave as capillary dipoles.  Dani \textit{et al.} \cite{Dani2015} conducted theoretical and experimental investigations of the capillary attractions of floating particles at an oil-water interface. The two-particle viscous drag coefficient was approximated analytically  and used to calculate the motion of two mutually approaching particles. Hemauer \textit{et al.} \cite{Hemauer2021} examined the capillary attractions between two freely rotating 2D particles at a deformable fluid interface through finite element implementations of a phase-field model. They  found that allowing the particles to rotate speeds up their capillary interactions.

 Despite much progress, an outstanding issue is  the impact of interfacial deformation on particle motion and in the presence of surface slip.
The objective of this study is to evaluate the effect of surface slip on the drag and torque exerted on a surface-trapped particle translating along an initially planar interface. We generalize the work of O'Neill \textit{et al.} \cite{oneill1986} to account for interface deformation.  Here we will only consider the case a non-rotating steadily translating particle, and hence there is no slip at the contact line for the steady-state cases we consider.  Following the two-parameter asymptotic modeling approach used in Ref. \cite{zhou2022}, we perturb the deformed interface from its planar state by assuming a small capillary number and a small contact angle deviation from 90 degrees. Our planar state solution is based on the work of  O'Neill \textit{et al.} \cite{oneill1986}. They solved for the flow past a  translating and rotating particle straddling a planar interface with prescribed Navier-slip boundary conditions on the particle surface. This is our leading-order problem. The leading order deformation due to the particle's translational motion is calculated analytically from the zeroth order stress found in Ref. \cite{oneill1986}. We derive explicit analytical expressions for the corresponding drag and torque coefficients of a translating sphere by applying the Lorentz reciprocal theorem to the zeroth and first-order approximations in our perturbation analysis. The drag and torque coefficients are calculated as functions of the three-phase contact angle, the Bond number (a dimensionless quantity measuring the relative effect of gravity forces and surface tension forces), and the slip coefficient. The asymptotic limits of large Bond number and zero surface slip are explored.  In addition, the  interaction forces and torques on a pair of translating and rotating particles are calculated via LSA.

\section{Mathematical model}

\begin{figure}
    \centering
    \includegraphics[width=15cm]{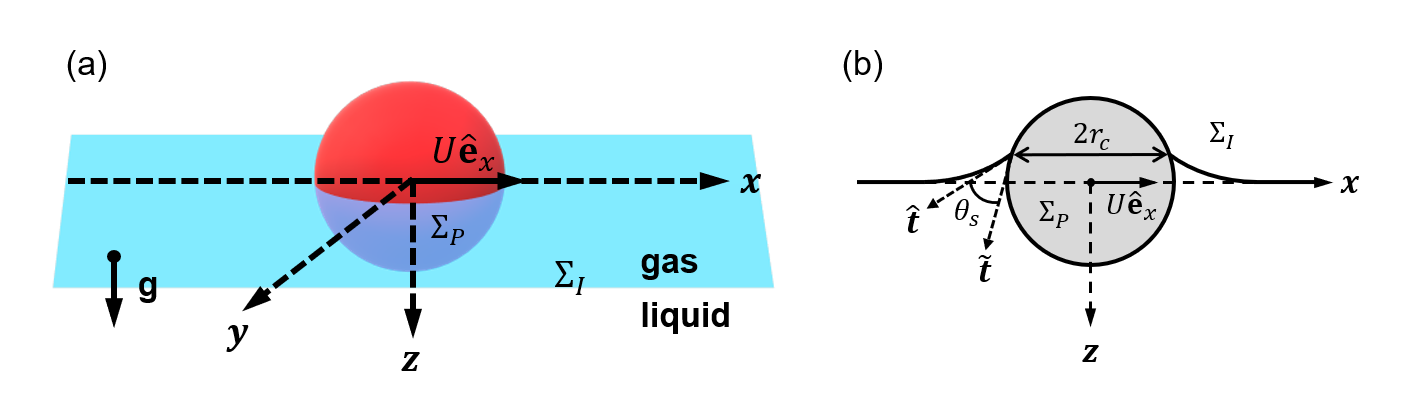}
    \caption{Illustration of (a) a spherical particle translating along a gas-liquid interface and (b) the cross-section view of a spherical particle translating along a deformed interface with contact angle $\theta_s.$ }
    \label{fig:sketch}
\end{figure}

Consider a translating  spherical particle of radius $a$ located midway at a  gas-liquid interface, as depicted in Fig.  \ref{fig:sketch}. 
We assume that the particle translates along the interface with velocity $ \mathbf{V} = U \hat{\mathbf{e}}_x$, and we are in the frame of reference moving with the translating particle, hence a steady solution can be expected.
In general this is a difficult free boundary problem for the fluid interface and for the location of the three-phase contact line along the particle interface.  Our approach is to make several simplifications to make the problems tractable to a perturbation analysis.  First, the Capillary number, $\mbox{Ca} = \mu U / \gamma$, 
is assumed small compared to unity implying that the interface should be near planar. Here, $\mu$ denotes the viscosity of the liquid phase, and $\gamma$ the interfacial tension.
Along the TCL, the contact angle $\theta_s$ is  constant and assumed close to $90^\circ$ so that the interfacial deformations remain small.
Gravitational forces (in the positive $z$ direction) are included in the model allowing us to define a Bond number $\text{Bo}  = \varrho g a^2/\gamma \sim 1.$ The density of the liquid is $\varrho$, and $g$ the gravitational acceleration.  
Finally we assume that the gaseous phase is  inviscid, of zero density, and of constant pressure.  In the following dimensionless variables are introduced with $U$ as the unit of velocity, the particle radius $a$ is the unit of length and  $\mu U/a$ is the unit of pressure.

For our analysis it is convient to introduce both spherical coordinates $(\rho, \theta,\varphi)$
\begin{align}
    x = \rho\sin\theta\cos\varphi,\quad y = \rho\sin\theta\sin\varphi,\quad z = \rho\cos\theta,
\end{align}
where $0 \leq \theta < \pi$ and  $0 \leq \varphi < 2 \pi$, and 
cylindrical coordinates $(r,\varphi,z)$,
\begin{align}
    x = r \cos\varphi,\quad y = r\sin\varphi. 
\end{align}
Here $\rho = 0$ is the particle center and $\rho = 1$ describes the particle surface.  We denote the gas-liquid interface as $z = h(r,\varphi)$ with  $r_c(\varphi) \leq r \leq \infty,$ where $r = r_c(\varphi)$ is the value of $r$ on the particle surface and satisfies the constraint $r_c^2(\varphi) + h^2(r_c,\phi) = 1.$ Let $\Sigma_P = \{ (r,\varphi,z) \vert  r^2 + z^2 =1, z \geq h(r_c,\varphi) \}$ denote the surface of the spherical particle immersed in the liquid phase.

The liquid flow is governed by the Stokes equations,
\begin{align}
    -\nabla p + \nabla^2\mathbf{u} & =\mathbf{0}, \\
    \nabla \cdot \mathbf{u} = 0.
    \end{align}
 At the particle surface, the traditional no-slip boundary conditions assume that the tangential velocities of the liquid at the solid surface equal the surface velocities. Here, the Navier-slip boundary conditions are adopted along the particle surface, 
\begin{align}
& \beta    \hat{\mathbf{e}}_\theta \cdot (\mathbf{u} - \hat{\mathbf{e}}_x) =  \hat{\mathbf{e}}_\rho \cdot \bm{\sigma}\cdot \hat{\mathbf{e}}_\theta, \label{eqn:slip_condition_theta}\\
&  \beta    \hat{\mathbf{e}}_\phi \cdot (\mathbf{u} - \hat{\mathbf{e}}_x) =  \hat{\mathbf{e}}_\rho \cdot \bm{\sigma}\cdot \hat{\mathbf{e}}_\phi, \quad \mbox{ at }\Sigma_P,  \label{eqn:slip_condition_phi}
\end{align}
where $\bm{\sigma} = -p\mathbf{I} + \nabla \mathbf{u} + (\nabla \mathbf{u})^T$ is the nondimensionalized stress tensor. The slip coefficient $\beta$ measures the tangential slip between the liquid and the solid surface of the particle. As $\beta \rightarrow \infty,$ the slip length becomes vanishingly small and the no-slip condition is recovered. In the case of flow past a gas bubble, $\beta = 0$ and the bubble's surface has perfect slip \cite{oneill1986}.  
The second condition on the velocity at the particle surface, the no-penetration condition,  requires that the normal velocity of the fluid equals to the normal velocity of the surface of the particle i.e., 
\begin{align}
   ( \mathbf{u}-\hat{\mathbf{e}}_x) \cdot \hat{\mathbf{e}}_\rho =0.\label{eqn:no-penetration_condition} 
\end{align}

Boundary conditions are still needed along the liquid interface.  Let $\Sigma_I = \{ (r,\varphi,z) \vert z = h(r,\varphi)  \}$ denote the liquid interface at steady state. At the interface the normal velocity and the shear stress vanish, i.e.,
\begin{align}
    & (\mathbf{u}-\hat{\mathbf{e}}_x) \cdot \hat{\mathbf{n}} = 0, \label{eqn:interface_condition1}\\
    & \hat{\mathbf{t}} \cdot \bm{\sigma} \cdot \hat{\mathbf{n}} = 0 \quad \mbox{ at }\Sigma_I, \label{eqn:interface_condition2}
\end{align}
where $\hat{\mathbf{n}}$ and $\hat{\mathbf{t}}$ are the respective unit normal and tangential vectors to the liquid interface $\Sigma_I$.
The normal stress is discontinuous across the liquid interface, and the discontinuity is balanced by interfacial tension and gravity effects, i.e.,
\begin{align}
 \hat{\mathbf{n}} 
 \cdot \bm{\sigma}\cdot \hat{\mathbf{n}} = \frac{1}{\mbox{Ca}} (\nabla\cdot\hat{\mathbf{n}}) +  \frac{\mbox{Bo}}{\mbox{Ca}} h . \label{eqn:stress_balance_condition}
\end{align}
In the far field, we require that the  fluid is at rest, which implies 
\begin{align}
    \mathbf{u}\rightarrow \mathbf{0} \quad \mbox{ as }\rho \rightarrow \infty. \label{eqn:far_field_condition}
\end{align}
Along the TCL, we assume that the contact angle $\theta_s$
 remains constant, where $\theta_s$ is defined to be the angle between the tangent to the liquid interface $\Sigma_I$ and the particle surface $\Sigma_P$ (see Fig. \ref{fig:sketch}b). This gives the contact angle condition,    
 \begin{align}
      \hat{\mathbf{n}} \cdot \hat{\mathbf{e}}_\rho = \cos\theta_s. \label{eqn:contact_angle_conditon}
 \end{align}

\section{Asymptotic expansions} \label{sec:asymptotic_expansions}

We perturb the deformed interface shape $z = h(r,\varphi)$ from the planar interface. Here, the interfacial deformation $h(r,\varphi)$ satisfies the normal stress balance equation \eqref{eqn:stress_balance_condition} and the contact angle condition \eqref{eqn:contact_angle_conditon}.
Adopting a similar asymptotic approach to that used in Ref. \cite{zhou2022}, we consider the two-parameter asymptotic expansion in the limit of small Ca and small deviation of the contact angle from $90^\circ$ for any quantity $f$, 
\begin{align}
    f = f^{(0,0)} + \mbox{Ca}f^{(1,0)} + \delta f^{(0,1)}  + \cdots, \label{eqn:asym_exp_f}
\end{align}
where Ca $\ll 1$ and $\theta_s = \pi/2+\delta \Tilde{\theta}_s$ with $\delta \ll1$ and $\Tilde{\theta}_s \sim 1.$ In addition, we assume $\text{Bo}\sim 1.$ The leading-order problem describes a translating sphere bisected by a gas-liquid interface, which is solved by O'Nell \textit{et al.} \cite{oneill1986} with slip boundary conditions \eqref{eqn:slip_condition_theta} and \eqref{eqn:slip_condition_phi}. Only the first correction beyond the leading order terms will be considered here. Higher order terms in this expansion would require a relation between $\text{Ca}$ and $\delta.$

The expansion of the deformed interface shape $h$ truncated after the $\mathcal{O}(\text{Ca})$ and $\mathcal{O}(\delta)$ terms is given by  
\begin{align}
    h \approx \text{Ca}h^{(1,0)}+  \delta h^{(0,1)}, \label{eqn:asym_exp_h}
\end{align}
where at the TCL, $r_c = 1 + \mathcal{O}(\text{Ca}^2 + \delta^2 + \text{Ca}\delta)$.
At $\mathcal{O}(1)$, the liquid interface remains flat, and the unperturbed interface and the particle surface immersed in the liquid phase are described by 
\begin{align}
   \Sigma_I^{(0,0)} = \{ (r,\varphi,z) \ | \  r\geq 1, z=0  \}, \quad   \Sigma_P^{(0,0)} = \{ (r,\varphi,z) \ | \ r^2 + z^2 = 1, z > 0 \},
\end{align}
respectively. 
At correction orders, the $\mathcal{O}(\text{Ca})$ deformation $\text{Ca} h^{(1,0)}$ accounts for the deformation induced by the particle's translational motion; the $\mathcal{O}(\delta)$ deformation $\delta h^{(0,1)}$ accounts for the deformation induced by the contact angle, which describes the steady-state interface shape around the particle independent of the particle motions.

\section{Interfacial deformations}\label{sec:interfacial_deformations}
Inserting the asymptotic expansion \eqref{eqn:asym_exp_h} into Eqs. \eqref{eqn:stress_balance_condition} and \eqref{eqn:contact_angle_conditon}
yields the linearized boundary value problems for the $\mathcal{O}(\delta)$ and $\mathcal{O}(\text{Ca})$ deformations. The static deformation $\delta h^{(0,1)}$ satisfies the $\mathcal{O}(\delta)$ stress balance equation  
 \begin{align}
    -   \frac{1}{r}\frac{\partial}{\partial r}\left( r\frac{\partial h^{(0,1)}}{\partial r}\right) - \frac{1}{r^2 } \frac{\partial^2 h^{(0,1)}}{\partial \varphi^2}+ \text{Bo} h^{(0,1)} =  0, \quad r \geq 1, \quad 0 \leq \varphi < 2\pi, 
 \end{align}
 with the  contact angle condition 
  \begin{align}
     -\tilde{\theta}_s = -\frac{\partial h^{(0,1)}}{\partial r} + h^{(0,1)}. \label{eqn:contact_angle_conditon_static_def}
 \end{align}
Additionally, we assume the static deformation vanishes in the far field, 
\begin{align}
    h^{(0,1)} \rightarrow 0\quad \text{ as } r \rightarrow \infty.  \label{eqn:far_field_condition_static_def}
\end{align}
 The analytical solution for $\delta h^{(0,1)}$ was obtained in our earlier work \cite{zhou2022}, 
\begin{align}
    \delta h^{(0,1)} = \delta (\tilde{b}-\Tilde{\theta_s}) C_0 K_0(\sqrt{\text{Bo}} r), \quad \text{with} \quad C_0 = \frac{1}{\sqrt{\text{Bo}} K_1(\sqrt{\text{Bo}}) + K_0(\sqrt{\text{Bo}})}, \label{eqn:static_deformation}
\end{align}
where  $K_n$ are the modified Bessel functions of order $n,$ and $b = \delta \tilde{b}$ denotes the particle's immersion depth into the liquid phase with the particle's center of mass located at $b \hat{\mathbf{e}}_z$.  Here, we assume the particle is located midway at the planar interface and thus the immersion depth $b = 0.$ 
In Appendix \ref{app:comparison_2D/3D_interfaces}, 
we compare the linear static deformation \eqref{eqn:static_deformation} with the numerical solutions of the 2D and 3D nonlinear interface shapes obtained by Loudet \textit{et al.} \cite{Loudet2020}  and Gudavadze and Florin \cite{Gudavadze2022}. The 3D linear result \eqref{eqn:static_deformation} demonstrates an excellent agreement with the 3D nonlinear solution for contact angle ranging from $75^\circ$ to $150^\circ$, while quantitative differences are observed between the 2D and 3D solutions. This implies that the model's dimensionality has a nontrivial impact on the equilibrium static deformation and that the asymptotic models in our earlier (Ref. \cite{zhou2022}) and current studies are robust to large variations in contact angle. Further numerical verification is required to understand the fully nonlinear effect of the contact angle.

 The correction deformation $h^{(1,0)}$ induced by the translational flow satisfies the normal stress balance equation \eqref{eqn:stress_balance_condition} at order Ca,
 \begin{align}
    -   \frac{1}{r}\frac{\partial}{\partial r}\left( r\frac{\partial h^{(1,0)}}{\partial r}\right) - \frac{1}{r^2 } \frac{\partial^2 h^{(1,0)}}{\partial \varphi^2}+ \text{Bo} h^{(1,0)} =  \sigma_{zz}^{(0,0)}, \quad r \geq 1, \quad 0 \leq \varphi < 2\pi, 
 \end{align}
 where the normal-normal stress $\sigma_{zz}^{(0,0)} = \hat{\mathbf{e}}_z \cdot \bm{\sigma}^{(0,0)} \cdot \hat{\mathbf{e}}_z$ at $z = 0$ can be computed from the zeroth-order flow fields obtained by O'Neill \textit{et al.} \cite{oneill1986}, i.e.,
 \begin{align}
      \sigma_{zz}^{(0,0)}  =  -\frac{3\beta}{2(3+\beta) r^4}\cos\varphi
 \end{align}

\begin{figure}
    \centering
    \includegraphics[width=15cm]{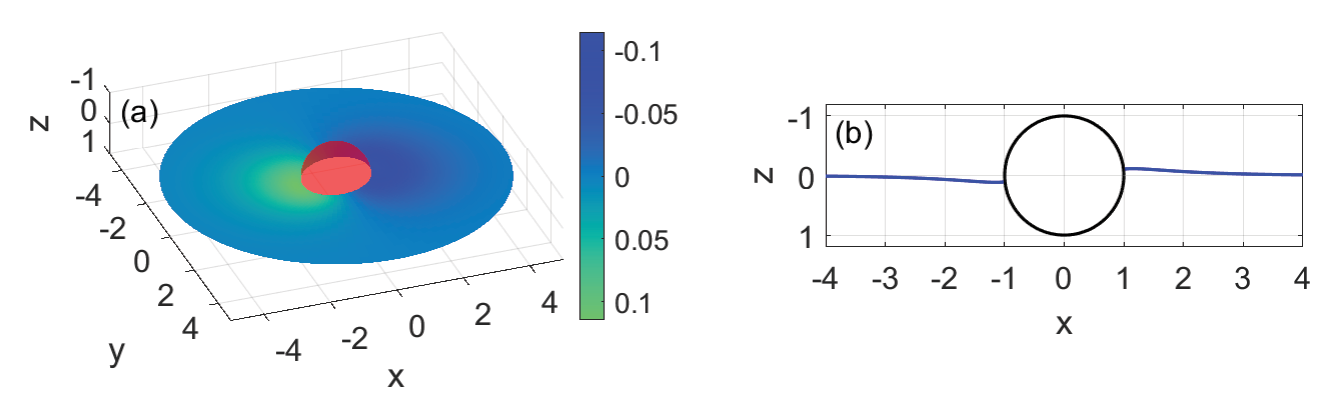}
    \caption{The flow-induced deformation $\text{Ca}h^{(1,0)}$ with $\text{Ca} = 1, \text{Bo} = 1,$ and $\beta = 10:$ (a) 3D visualization, where the color map shows the interfacial height around the particle; (b) $x$-$z$ cross-section view. }
\label{fig:deformation_3d}
\end{figure}

At the TCL and in the far field, $h^{(1,0)}$ satisfy the boundary conditions
 \begin{align}
    &  - \frac{\partial h^{(1,0)}}{\partial r} + h^{(1,0)} =0 \quad \text{at } r = 1,  \\
   &  h^{(1,0)} \rightarrow 0 \quad \text{ as } r \rightarrow \infty. 
 \end{align}
 The solution for $h^{(1,0)}$ can be obtained using the method of variation of parameters, which is given by
 \begin{align}
 \begin{split}
   h^{(1,0)} = \left[C_1 K_1(\sqrt{\mbox{Bo}} r )   +\right. &  K_1(\sqrt{\mbox{Bo}} r ) \int_1^r \frac{-3\beta}{2(3+\beta) r^3} I_1(\sqrt{\mbox{Bo}} r ) \mbox{ d}r \\
    + &  \left. I_1(\sqrt{\mbox{Bo}} r ) \int_r^\infty  \frac{-3\beta}{2(3+\beta) r^3} K_1(\sqrt{\mbox{Bo}} r ) \mbox{ d}r\right]\cos\varphi,
\end{split}\label{eqn:flow_induced_deformations_solutions}
 \end{align}
 with 
 \begin{align}
     C_1 = \frac{I_2(\sqrt{\text{Bo}}) \int_1^\infty \frac{-3\beta}{2(3+\beta) r^3}K_1(\sqrt{\text{Bo}}r)\mbox{ d}r}{K_2(\sqrt{\text{Bo}})},
 \end{align}
where $K_n$ and $I_n$ are the modified Bessel functions of order $n.$
The flow-induced deformation $h^{(1,0)}$ is visualized in Fig. \ref{fig:deformation_3d} with $\text{Ca} = 1, \text{Bo} = 1$, and $\beta = 10.$ 

\begin{figure}
    \centering
    \includegraphics[scale=0.6]{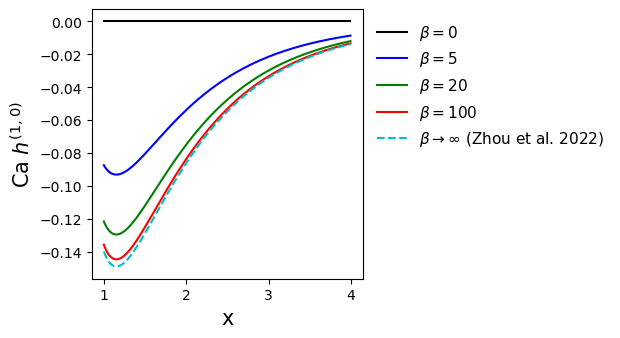}
    \caption{ The flow-induced deformation $\text{Ca}  h^{(1,0)}$ plotted along the positive $x$-axis with $\text{Ca} =1, \text{Bo}=1$, and varying values of slip coefficient $\beta.$ }
    \label{fig:deformations2}
\end{figure}

In Fig.  \ref{fig:deformations2}, we show the interfacial profile of Ca$h^{(1,0)}$ along the positive $x$-axis with Ca =1, Bo = 1, and varying values of slip coefficient $\beta.$ Figure \ref{fig:deformations2} shows that when the particle surface has a perfect slip ($\beta = 0$), the interface shape is unaffected by the particle motions and $h^{(1,0)} \equiv 0.$ As $\beta$ increases, the slip length decreases, and the amplitude of the interfacial deformation increases. As $\beta$ becomes infinitely large, the flow-induced deformation $h^{(1,0)}$ approaches the no-slip solution obtained by Ref. \cite{zhou2022} as shown in Fig. \ref{fig:deformations2}.

\begin{figure}
    \centering
    \includegraphics[scale=0.6]{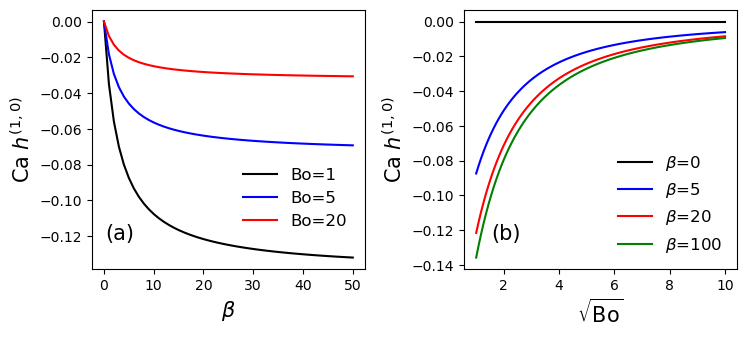}
    \caption{The interfacial height of $\text{Ca} h^{(1,0)}$ at $(r,\varphi) = (1,0)$ plotted as functions of (a) slip coefficient $\beta$ with varying values of Bond number Bo, and (b) $\sqrt{\text{Bo}}$ with varying values of slip coefficient $\beta$, with $\text{Ca} = 1$.  }
    \label{fig:deformations3}
\end{figure}

    To further examine how the slip coefficient affects the flow-induced deformation, in Fig.  \ref{fig:deformations3}(a), we plot the interface height of $\text{Ca} h^{(1,0)}$ at $(r,\varphi) = (1,0)$ as a function of the slip coefficient $\beta$ with $\text{Ca} = 1$ and varying values of Bond number. Figure \ref{fig:deformations3}(a) shows that, at the TCL, the elevation of the interface deformed by the particle's translational motion remains bounded and converges to a finite value as $\beta\rightarrow\infty$.  In Fig. \ref{fig:deformations3}(b), we plot the interfacial height $\text{Ca}h^{(1,0)}$ at $(r,\varphi) = (1,0)$ as a function of $\sqrt{\text{Bo}}$ for $\beta = 0, 5, 20,$ and 100. Recall that Bond number characterizes the density of the liquid phase. As $\sqrt{\text{Bo}}$ increases, the magnitude of the interfacial height decreases due to the increased gravity force acting on the interface, as shown in Fig. \ref{fig:deformations3}(b). 

\section{Drag and torque on a translating sphere}
\label{sec:trans:translation_drag_torque}
The nondimensionalized  hydrodynamic drag and torque exerted on a spherical particle with translation velocity $\mathbf{V} 
= \hat{\mathbf{e}}_x$ are given, respectively, by 
\begin{align}
 F &  = \iint_{\Sigma_P} (  \bm{\sigma} \cdot \hat{\mathbf{e}}_\rho ) \cdot \hat{\mathbf{e}}_x \text{ d}\Sigma, \label{eqn:trans:F_t_formula} \\
  T  &   = \iint_{\Sigma_P}  \mathbf{x}_P \times (\bm{\sigma} \cdot \hat{\mathbf{e}}_\rho )\cdot \hat{\mathbf{e}}_y \text{ d}\Sigma,\label{eqn:trans:T_t_formula1}
\end{align}
where $\mathbf{x}_P = \hat{\mathbf{e}}_\rho$ is the position vector on the particle surface.
Substituting the expansions \eqref{eqn:asym_exp_f} and \eqref{eqn:asym_exp_h} for $\bm{\sigma}$ and $h$ into Eq. \eqref{eqn:trans:F_t_formula}, we can express $F$ as an asymptotic series to orders Ca and $\delta$,
\begin{align}
\begin{split}
    F
    = &  \iint_{\Sigma_P^{(0,0)}} \left(\bm{\sigma}^{t(0,0)} + \mbox{Ca} \bm{\sigma}^{(1,0)} + \delta \bm{\sigma}^{(0,1)} \right) \cdot \hat{\mathbf{e}}_\rho\cdot \hat{\mathbf{e}}_x  \mbox{ d}\Sigma\\
    &- \int_{\Sigma_{\text{\tiny TCL}}^{(0,0)}}   \left( \mbox{Ca}  h^{(1,0)}  + \delta  h^{(0,1)}\right)  \bm{\sigma}^{(0,0)}\cdot  \hat{\mathbf{e}}_r \cdot \hat{\mathbf{e}}_x  \mbox{ d}s \\
   = &   F^{(0,0)} + \text{Ca} F^{(1,0)} + \delta F^{(0,1)},
   \end{split}\label{eqn:trans:F_t_formula2} 
\end{align}
where $\Sigma_{\text{\tiny TCL}}^{(0,0)} = \{ (r,\varphi,z) \vert r=1, z=0\}.$
Applying the vector identity $\mathbf{b} \times \mathbf{c} \cdot \mathbf{a} =\mathbf{a}\times\mathbf{b}\cdot\mathbf{c}$ to Eq. \eqref{eqn:trans:T_t_formula1}, the hydrodynamic torque can be rewritten as 
\begin{align}
    T = \iint_{\Sigma_P} \hat{\mathbf{e}}_y \times \hat{\mathbf{e}}_\rho \cdot  (\bm{\sigma} \cdot \hat{\mathbf{e}}_\rho )  \text{ d}\Sigma =  \iint_{\Sigma_P} (\bm{\sigma} \cdot \hat{\mathbf{e}}_\rho ) \cdot (\hat{\mathbf{e}}_y
    \times \hat{\mathbf{e}}_\rho) \text{ d}\Sigma. \label{eqn:trans:T_t_formula2}
\end{align}
Inserting the expansion \eqref{eqn:asym_exp_f} into Eq. \eqref{eqn:trans:T_t_formula2}, we obtain a truncated asymptotic series for $T$, 
\begin{align}
\begin{split}
    T    = & \iint_{\Sigma_P^{(0,0)}}\left(\bm{\sigma}^{(0,0)} + \mbox{Ca} \bm{\sigma}^{(1,0)} + \delta \bm{\sigma}^{(0,1)} \right) \cdot \hat{\mathbf{e}}_\rho\cdot (\hat{\mathbf{e}}_y
    \times \hat{\mathbf{e}}_\rho)   \mbox{ d}z  \mbox{ d}\varphi\\
    &- \int_{\Sigma_{\text{\tiny TCL}}^{(0,0)}}  \left( \mbox{Ca}  h^{(1,0)}  + \delta  h^{(0,1)}\right)  \bm{\sigma}^{(0,0)}\cdot  \hat{\mathbf{e}}_r \cdot (\hat{\mathbf{e}}_y
    \times \hat{\mathbf{e}}_\rho)   \mbox{ d}s \\
   = &   T^{(0,0)} + \text{Ca} T^{(1,0)} + \delta T^{(0,1)}.  
\end{split}\label{eqn:trans:T_t_formula3}
\end{align}
O'Neill \textit{et al.} \cite{oneill1986} obtained the $\mathcal{O}(1)$ drag and torque acting on a translating particle, i.e., 
\begin{align}
     F^{(0,0)} = -k_D^{(0,0)} =  -\frac{3\pi (\beta+2)}{\beta+3}, \quad T^{(0,0)} = -k_T^{(0,0)} = -\frac{3\pi \beta }{2(\beta+3)},\label{eqn:trans:O(1)drag_torque_solutions}
\end{align}
where $k_D^{(0,0)}$ and $k_T^{(0,0)}$ are the respective drag and torque coefficients. 
The $\mathcal{O}(\text{Ca})$ and $\mathcal{O}(\delta)$ drag and torque can be obtained by following the analytical approach described in Ref. \cite{zhou2022} by applying the Lorentz reciprocal theorem. 

\subsection{Drag coefficient of a translating sphere} \label{subsec:trans:drag_calculation}
At orders Ca and $\delta$, $F^{(1,0)}$ represents the correction drag due to the flow-induced deformation $\text{Ca} h^{(1,0)}$, and $F^{(0,1)}$ denotes the contact-angle-induced correction drag due to the static deformation $\delta h^{(1,0)}$. The formula of the correction drags are given by 
\begin{align}
    F^{(j,k)} = \underbrace{\iint_{\Sigma_P^{(0,0)}} \bm{\sigma}^{(j,k)} \cdot \hat{\mathbf{e}}_\rho \cdot \hat{\mathbf{e}}_x  \text{ d}\Sigma}_{\text{\textcircled{\scriptsize 1}}} \underbrace{ - \int_{\Sigma_{\text{\tiny TCL}}^{(0,0)}} h^{(j,k)} \bm{\sigma}^{(0,0)} \cdot \hat{\mathbf{e}}_r \cdot \hat{\mathbf{e}}_x  \text{ d}s  }_{\text{\textcircled{\scriptsize 2}}},  \label{eqn:trans:corr_drag_formula}
 \end{align}
with $(j,k) = (1,0)$ and $(0,1)$. Recall that the static deformation 
in Eq. \eqref{eqn:static_deformation} is independent of the particle motion. 
In Eq. \eqref{eqn:trans:corr_drag_formula}, the integral \text{\textcircled{\small 2}} accounts for the drag contribution from the TCL's deviation from the unit circle $r = 1$ and can be directly evaluated.  For $(j,k) = (1,0), \text{\textcircled{\small 2}} = 0$, and for $(j,k) = (0,1),$
\begin{align}
    \text{\textcircled{\small 2}} = \int_0^{2\pi}   h^{(0,1)}(1) \left( \frac{3}{2} + \frac{9 \cos2\varphi}{2(3+\beta)} \right)\text{ d}\varphi = 3\pi h^{(0,1)}(1)=-3 \pi \Tilde{\theta}_s C_0 K_0(\sqrt{\text{Bo}} ) . \label{eqn:trans:integral_2}
\end{align}
It appears that the drag contribution from the TCL is unaffected by the inclusion of slip, and Eq. \eqref{eqn:trans:integral_2} recovers the last term in Eq. (49) in Ref. \cite{zhou2022}, where a no-slip model is considered.  

The term \textcircled{\small 1} in Eq. \eqref{eqn:trans:corr_drag_formula} can be calculated by using the Lorentz reciprocal theorem. We consider two Stokes flow problems defined in the same geometry of a semi-infinite fluid bounded by the particle surface  $\Sigma_P^{(0,0)}$, the free surface $\Sigma_I^{(0,0)}$, and the hemispherical surface $\Sigma_\infty$ with infinite radius. The first problem is the $\mathcal{O}(1)$ problem that describes a translating sphere with surface velocity $\hat{\mathbf{e}}_x$ bisected by an undeformed gas-liquid interface with velocity field $\mathbf{u}^{(0,0)}$ and stress $\bm{\sigma}^{(0,0)}$. 
At the particle surface $\Sigma_P^{(0,0)}$, $\mathbf{u}^{(0,0)}$ and $\bm{\sigma}^{(0,0)}$ satisfy the $\mathcal{O}(1)$ Navier-slip conditions \eqref{eqn:slip_condition_theta} and \eqref{eqn:slip_condition_phi} and no-penetration condition \eqref{eqn:no-penetration_condition}.
This implies \cite{phdApril}
\begin{align}
    \mathbf{u}^{(0,0)} = 
    V_\rho \hat{\mathbf{e}}_\rho + \left(V_\theta + \frac{1}{\beta} \sigma^{(0,0)}_{\rho\theta}\right) \hat{\mathbf{e}}_\theta  +  \left(V_\phi + \frac{1}{\beta} \sigma^{(0,0)}_{\rho\phi}\right)\hat{\mathbf{e}}_\phi, \label{eqn:trans:reciprocal_slip_condition_O(1)}
\end{align}
where $\hat{\mathbf{e}}_x =V_\rho \hat{\mathbf{e}}_\rho + V_\theta \hat{\mathbf{e}}_\theta + V_\phi\hat{\mathbf{e}}_\phi = \sin\theta\cos\phi\hat{\mathbf{e}}_\rho + \cos\theta\cos\phi\hat{\mathbf{e}}_\theta -\sin\phi\hat{\mathbf{e}}_\phi.  $
The $\mathcal{O}(1)$ interface conditions \eqref{eqn:interface_condition1} and \eqref{eqn:interface_condition2} and far-field condition \eqref{eqn:far_field_condition} yield
\begin{align}
    \mathbf{u}^{(0,0)} \cdot \hat{\mathbf{e}}_z = 0,\quad \hat{\mathbf{t}}^{(0,0)} \cdot \bm{\sigma}^{(0,0)} \cdot \hat{\mathbf{e}}_z = 0,
    \end{align}
    at $\Sigma_I^{(0,0)}$, and $\mathbf{u}^{(0,0)} = 0$ at $\Sigma_\infty$.

To use the Lorentz reciprocal theorem, the second problem is constructed by the truncated asymptotic expansions of the velocity 
 field $\mathbf{u}$ and stress $\bm{\sigma} $: 
 \begin{align}
     \mathbf{u}  & = \mathbf{u}^{(0,0)} + \text{Ca} \mathbf{u}^{(1,0)} + \delta \mathbf{u}^{(0,1)},\label{eqn:trans:asym_exp_for_vel} \\
     \bm{\sigma} & = \bm{\sigma}^{(0,0)} + \text{Ca} \bm{\sigma}^{(1,0)} + \delta \bm{\sigma}^{(0,1)}. \label{eqn:trans:asym_exp_for_stress}
 \end{align}
The $\mathcal{O}(\text{Ca})$ and $\mathcal{O}(\delta)$ boundary and interface conditions \eqref{eqn:slip_condition_theta} - \eqref{eqn:far_field_condition} require 
\begin{align}
  \mathbf{u}^{(j,k)} =  & \frac{1}{\beta} \left( \hat{\mathbf{e}}_\rho \cdot \bm{\sigma}^{(j,k)} \cdot \hat{\mathbf{e}}_\theta  + \hat{\mathbf{e}}_\rho \cdot \bm{\sigma}^{(j,k)} \cdot \hat{\mathbf{e}}_\phi\right)=   \frac{1}{\beta} \sigma^{(j,k)}_{\rho\theta} \hat{\mathbf{e}}_\theta  +   \frac{1}{\beta} \sigma^{(j,k)}_{\rho\phi} \hat{\mathbf{e}}_\phi, \label{eqn:trans:reciprocal_slip_condition}
\end{align}
at the particle surface $\Sigma_P^{(0,0)}$,
\begin{align}
     &  \mathbf{u}^{(j,k)} \cdot \hat{\mathbf{e}}_z = -(\mathbf{u}^{(0,0)}-\hat{\mathbf{e}}_x)\cdot \hat{\mathbf{n}}^{(j,k)} - \frac{\partial \mathbf{u}^{(0,0)}}{\partial z} h^{(j,k)}\cdot \hat{\mathbf{e}}_z ,\label{eqn:trans:reciprocal:interface_condition1}\\ 
  &    \hat{\mathbf{t}}^{(0,0)} \cdot  \bm{\sigma}^{(j,k)} \cdot \hat{\mathbf{e}}_z =  - \hat{\mathbf{t}}^{(0,0)} \cdot  \frac{\partial \bm{\sigma}^{(0,0)}}{\partial z }h^{(j,k)} \cdot \hat{\mathbf{e}}_z - \hat{\mathbf{t}}^{(0,0)} \cdot \bm{\sigma}^{(0,0)} \cdot \hat{\mathbf{n}}^{(j,k)} - \hat{\mathbf{t}}^{(j,k)} \cdot \bm{\sigma}^{(0,0)} \cdot \hat{\mathbf{e}}_z,\label{eqn:trans:reciprocal:interface_condition2}
\end{align}
at the interface $\Sigma_I^{(0,0)}$, and $\mathbf{u}^{(j,k)} = 0$ at $\Sigma_\infty$, for $(j,k) = (1,0)$ and $(0,1)$.

The two Stokes flow problems are related by the Lorentz reciprocal theorem (\cite{Happel-1983,Lorentz-1896})
\begin{align}
    \iint\limits_{\Sigma_P^{(0,0)} \cup\Sigma_I^{(0,0)} \cup \Sigma_\infty} \bm{\sigma}^{(0,0)} \cdot \mathbf{n} \cdot \mathbf{u} \text{ d}\Sigma =\iint\limits_{\Sigma_P^{(0,0)} \cup\Sigma_I^{(0,0)} \cup \Sigma_\infty} \bm{\sigma}^{} \cdot \mathbf{n} \cdot \mathbf{u}^{(0,0)} \text{ d}\Sigma, \label{eqn:trans:reciprocal_theorem_relation}
\end{align} 
where $\mathbf{n}$ is the unit normal to the surface bounding the semi-infinite fluid.
Inserting Eqs.  \eqref{eqn:trans:asym_exp_for_vel} - \eqref{eqn:trans:reciprocal_slip_condition} into Eq. \eqref{eqn:trans:reciprocal_theorem_relation} and rearranging terms, we obtain  
\begin{align}
 &   
\text{\textcircled{\small 1}} = \iint_{\Sigma_{P}^{(0,0)}} \bm{\sigma}^{(j,k)} \cdot \hat{\mathbf{e}}_\rho  \cdot \mathbf{V}^t  \mbox{ d}\Sigma = \iint_{\Sigma_I^{(0,0)}} \bm{\sigma}^{(0,0)}\cdot \hat{\mathbf{e}}_z\cdot \mathbf{u}^{(j,k)} \mbox{ d}\Sigma -  \iint_{\Sigma_I^{(0,0)}} \bm{\sigma}^{(j,k)}\cdot \hat{\mathbf{e}}_z\cdot \mathbf{u}^{(0,0)} \mbox{ d}\Sigma,\label{eqn:trans:reciprocal_eqn1}
\end{align}
with $(j,k) = (1,0)$ and $(0,1)$ (see Appendix \ref{app:reciprocal_theorem:drag} for derivation details). Note that after substituting Eqs. \eqref{eqn:trans:reciprocal:interface_condition1} and \eqref{eqn:trans:reciprocal:interface_condition2} 
into Eq. \eqref{eqn:trans:reciprocal_eqn1}, the integral \textcircled{\small 1} can be expressed solely in terms of the $\mathcal{O}(1)$ solutions, which are given by Eqs. \eqref{eqn:app:order1_sol1} - \eqref{eqn:app:order1_sol4}, and the $\mathcal{O}(\text{Ca})$ and $\mathcal{O}(\delta)$ deformations. The correction drags $F^{(1,0)}$ and $F^{(0,1)}$ can now be  evaluated. Because of the antisymmetric configuration of the flow-induced deformation $\text{Ca} h^{(1,0)}$, the $\mathcal{O}(\text{Ca})$ stress at the front and back of a translating particle cancel each other out and the $\mathcal{O}(\text{Ca})$ drag contribution vanishes, i,e, $F^{(1,0)} = 0$. The $\mathcal{O}(\delta)$ correction drag, $F^{(0,1)}$, due to the static deformation $\delta h^{(0,1)}$ is derived to be
{\small 
\begin{align}
\begin{split}
 F^{(0,1)}  =\int_1^\infty  & \frac{3\pi}{8 r^7 (3+\beta)^2} \left\{  -6\left[ 2\beta^2 - 5 r^2 \beta(2+\beta) + 3 r^4(2+\beta)^2 \right]h^{(0,1)} \right.\\
  & + \left. r \left[ 9\beta^2-27r^2\beta(2+\beta) + 18 r^4 (2+\beta)^2 + 4 r^3\beta(3+\beta) \right]\frac{\partial h^{(0,1)}}{\partial r}\right\} \mbox{ d}r \\ & -  3\pi \Tilde{\theta}_s C_0 K_0(\sqrt{\text{Bo}}).
   \end{split}  \label{eqn:trans:orderDeltaDrag} 
\end{align}}

We define the $\mathcal{O}(\delta)$ drag coefficient $k^{(0,1)}_D(\text{Bo},\beta)$  by the relation $F^{(1,0)}=-\Tilde{\theta}_s k^{(0,1)}_D$, where $k^{(0,1)}_D$ can be expanded in factors in terms of $\beta$: 

\begin{align}
    k_D^{(0,1)} = \frac{1}{(\beta+3)^2} \left(\mathcal{I}_0 +\beta \mathcal{I}_1  + \beta^2 \mathcal{I}_2  \right) + 3\pi C_0 K_0(\sqrt{\text{Bo}}), \label{eqn:trans_kDt_beta_exp}
\end{align}
where $\mathcal{I}_0, \mathcal{I}_1$, and $\mathcal{I}_2$ given by
\begin{align}
    \mathcal{I}_0 &  = -27 \pi C_0 \int_1^\infty  \frac{1}{r^3}  \left( K_0(\sqrt{\text{Bo}} r )   +  \sqrt{\text{Bo} } r K_1(\sqrt{\text{Bo}}  r )   \right)\text{ d}r, \label{eqn:trans:I_0}\\
    \mathcal{I}_1 &  = 9 \pi  C_0 \int_1^\infty \frac{1}{4 r^5} \left[ 2 (5-6r^2) K_0(\sqrt{\text{Bo}} r)  +  \sqrt{\text{Bo}} r (9-2r(1+6r)) K_1(\sqrt{\text{Bo}} r ) \right] \text{ d}r \label{eqn:trans:I_1}\\ 
    \mathcal{I}_2 & =  -3 \pi C_0 \int_1^\infty \frac{1}{8 r^7} \left[  6(2-5r^2+3r^4) K_0(\sqrt{\text{Bo}} r ) \right. \nonumber \\  & +  \left. \sqrt{\text{Bo}} r(9+r^2(-27+2r(2+9r))) K_1(\sqrt{\text{Bo}} r ) \right]\text{ d}r.\label{eqn:trans:I_2}
\end{align}

Note that the $\beta$-dependence is explicit and only in the first term of Eq. \eqref{eqn:trans_kDt_beta_exp}. To leading order as $\beta \rightarrow\infty,$ 
\begin{align}
k^{(0,1)}_{D,\infty} = \mathcal{I}_2 +3\pi C_0 K_0(\sqrt{\text{Bo}})    
\end{align}
recovers the $\mathcal{O}(\delta)$ drag coefficient in the no-slip model in our earlier work \cite{zhou2022}. 
In the other limiting case, when the slip on the particle surface approaches a perfect slip ($\beta \rightarrow 0$), we obtain 
\begin{align}
    k^{(0,1)}_{D,0} = \frac{1}{9}\mathcal{I}_0 +3\pi C_0 K_0(\sqrt{\text{Bo}}).\label{eqn:trans:perfect_slip_limit}
\end{align}

\begin{figure}
    \centering
    \includegraphics[scale=0.6]{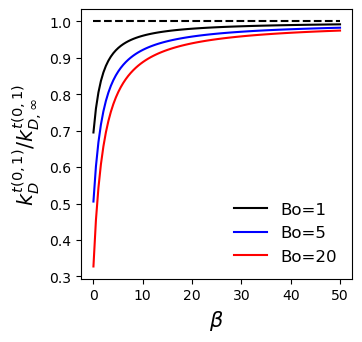}
    \caption{  The normalized drag coefficient  $k^{(0,1)}_{D,\infty}/k^{(0,1)}_{D,\infty}$ plotted as a function of slip coefficient $\beta$ with $\text{Bo} = 1, 5,$ and 20.}
    \label{fig:trans_drag_slip_conv_slip}
\end{figure}

\begin{figure}
    \centering
    \includegraphics[scale=0.6]{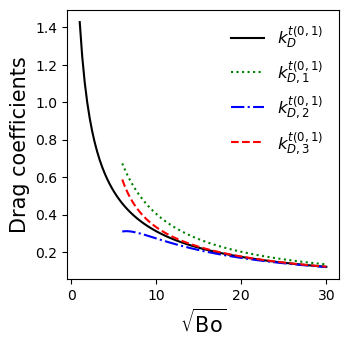}
    \caption{The drag coefficient $k^{(0,1)}_D$ and its large-Bo asymptotic approximations $k^{(0,1)}_{D,n}$ plotted as a functions of $\sqrt{\text{Bo}}$ with $\beta =10. $ }
    \label{fig:trans_drag_q_asym}
\end{figure}

The drag coefficient $k^{(0,1)}$ is computed by numerically integrating Eqs.  \eqref{eqn:trans:I_0} - \eqref{eqn:trans:I_2} using an adaptive quadrature from the Fortran library QUADPACK.  In Fig. \ref{fig:trans_drag_slip_conv_slip}, we plot the drag coefficient $k^{(0,1)}_D$ normalized by its no-slip limit,  $k^{(0,1)}_D/k^{(0,1)}_{D,\infty}$, as a function of $\beta.$ Figure \ref{fig:trans_drag_slip_conv_slip} shows that as $\beta$ increases from zero, $k^{(0,1)}_D$ increases and asymptotes to the no-slip drag coefficient $k^{(0,1)}_{D,\infty}$, meaning a larger surface slip (smaller slip coefficient $\beta$) on the particle reduces the correction drag $\delta F^{(0,1)}$ exerted on the particle due to the static deformation $\delta h^{(0,1)}$. 
In Fig.\ref{fig:trans_drag_q_asym}, the black curve shows the drag coefficient $k^{(0,1)}$ as a function of the Bond number with $\beta = 10$.  We observe that, in the slip model, the correction drag's dependency on Bo is similar to that in the no-slip model in Ref. \cite{zhou2022}, that is, as the Bond number increases, the increased liquid density flattens the interface and reduces correction drag $\delta F^{(0,1)}$. In the limit of large Bond number ($\text{Bo} \rightarrow\infty$), the asymptotic behavior of $k^{(0,1)}_D$ is described by 
{\small 
\begin{align}
    k^{(0,1)}_D = \frac{3\pi \beta(9+2\beta) }{4(3+\beta)^2 \sqrt{\text{Bo}}} -  \frac{9\pi(-24+25\beta + 4\beta^2)}{8(3+\beta)^2 \text{Bo}} + \frac{3\pi (-189+630\beta+25\beta^2)}{8(3+\beta)^2 \text{Bo}^{3/2}} + \mathcal{O}\left(\frac{1}{\text{Bo}^2}\right). \label{eqn:trans:large-Bo_drag_approx}
\end{align}
}
The large-Bo approximations of  $k^{(0,1)}_D$ are plotted in Fig. \ref{fig:trans_drag_q_asym} along with the exact solution, where $k^{(0,1)}_{D,n}$ denotes the asymptotic series in Eq. \eqref{eqn:trans:large-Bo_drag_approx} truncated after the $\text{Bo}^{-n/2}$ term. In the limit of Bo $\rightarrow 0$, the magnitude of the static deformation $h^{(0,1)}$ far away from the particle decays to zero at a much slower rate and fails to satisfy the far-field condition \eqref{eqn:far_field_condition_static_def} for $\text{Bo} = 0$ (see details in Appendix \ref{App:asymptitics_bond_number} and Ref. \cite{zhou2022}).  As a result, the correction drag $\delta F^{(0,1)}$ due to the static deformation becomes invalid at $\text{Bo} = 0.$

\subsection{Torque coefficient of a translating sphere}\label{subsec:trans:torque_calculation}
The orders Ca and $\delta$ hydrodynamic torques acting on a translating particle from Eq. \eqref{eqn:trans:T_t_formula2} are given by 
\begin{align}
    T^{(j,k)} = \underbrace{\iint_{\Sigma_P^{(0,0)}} \bm{\sigma}^{(j,k)} \cdot \hat{\mathbf{e}}_\rho \cdot (\hat{\mathbf{e}}_y\times\hat{\mathbf{e}}_\rho)  \text{ d}\Sigma}_{\text{\textcircled{\scriptsize 3}}} \underbrace{ - \int_0^{2\pi} h^{(j,k)} \bm{\sigma}^{(0,0)} \cdot \hat{\mathbf{e}}_r \cdot (\hat{\mathbf{e}}_y\times\hat{\mathbf{e}}_\rho)  \text{ d}\varphi }_{\text{\textcircled{\scriptsize 4}}},  \label{eqn:trans:corr_torque_formula}
 \end{align}
 with $(j,k) = (1,0)$ and $(0,1)$. The integral \textcircled{\small 4} can be evaluated analytically and equals to zero for both $T^{(1,0)}$ and  $T^{(0,1)}$.
Similar to the drag force calculation in Sec. \ref{subsec:trans:drag_calculation}, the integral  
 \textcircled{\small 3} can be evaluated by applying the Lorentz reciprocal theorem.  We consider two Stokes flow problems that are defined in the same domain bounded by $\Sigma_P^{(0,0)}, \Sigma_I^{(0,0)},$ and $\Sigma_\infty$ and therefore connected by the Lorentz reciprocal relation. 

The first problem describes a rotating spherical particle with surface slip bisected by an undeformable planar interface. It is constructed by prescribing the rotational velocity $\mathbf{V}^r = \hat{\mathbf{e}}_y \times \hat{\mathbf{e}}_\rho$ on the particle surface and setting Ca and $\delta$ to zero. We let $\mathbf{u}^r$ and $\bm{\sigma}^r$ denote the respective velocity field and stress induced by the rotational motion of the particle. At the particle surface, $\mathbf{u}^{r(0,0)}$ and $\bm{\sigma}^{r(0,0)}$ satisfy the Navier-slip and no-penetration conditions,
\begin{align}
       ( \mathbf{u}^{r(0,0)} - \hat{\mathbf{e}}_y\times\hat{\mathbf{e}}_\rho) 
   \cdot  \hat{\mathbf{e}}_\theta &  = \frac{1}{\beta}\hat{\mathbf{e}}_\rho \cdot \bm{\sigma}^{r(0,0)} \cdot \hat{\mathbf{e}}_\theta, \label{eqn:rot:navier_slip_cond_theta}\\
      (  \mathbf{u}^{r(0,0)} - \hat{\mathbf{e}}_y\times\hat{\mathbf{e}}_\rho) \cdot \hat{\mathbf{e}}_\varphi & = \frac{1}{\beta}\hat{\mathbf{e}}_\rho \cdot \bm{\sigma}^{r(0,0)} \cdot \hat{\mathbf{e}}_\phi, \label{eqn:rot:navier_slip_cond_phi}\\
       (\mathbf{u}^{r(0,0)} - \hat{\mathbf{e}}_y\times\hat{\mathbf{e}}_\rho) \cdot \hat{\mathbf{e}}_\rho &  = 0.\label{eqn:rot:no_slip_cond}\\
\end{align}
At the planar interface $\Sigma_I^{(0,0)}$, the normal velocity and the tangential stress vanishes, 
\begin{align}
    \mathbf{u}^{r(0,0)} \cdot \hat{\mathbf{e}}_z = 0,\quad \hat{\mathbf{t}}^{(0,0)} \cdot \bm{\sigma}^{r(0,0)} \cdot \hat{\mathbf{e }}_z = 0. 
\end{align}
In the far field, we require that $\mathbf{u}^{r(0,0)}  = 0$ at $\Sigma_\infty$.
The analytical solution of the rotational flow field $\mathbf{u}^{r(0,0)}$ is obtained by O'Neill \textit{et al.} \cite{oneill1986} through an eigenfunction expansion in Legendre polynomials (more details in Appendix \ref{App:order1_solution}).  

\begin{figure}
    \centering
    \includegraphics[scale=0.6]{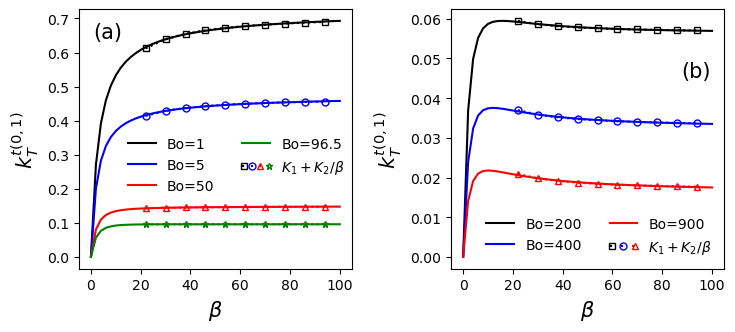}
    \caption{The torque coefficient $k_T^{(0,1)}$ plotted as a function of slip coefficient $\beta$ in comparison with their large-$\beta$ asymptotic approximations $K_1 + K_2/\beta$ with (a) Bo = 1, 5, 50, and 96.5 and (b) Bo = 200, 400, and 900. For $\text{Bo} = 1$, $K_1 \approx 0.71$ and $K_2 \approx -2.23$, for $\text{Bo} = 5, $ $K_1 \approx 0.47$ and $K_2 \approx -1.21$, for $\text{Bo} = 50$ $K_1 \approx 0.15$ and $K_2 \approx -0.14$, and for $\text{Bo} = 96.5$, $K_1 \approx 0.1$ and $K_2 \approx -0.01$;  for $\text{Bo} = 200$, $K_1 \approx 0.056$ and $K_2 \approx 0.081$,  for $\text{Bo} = 400$, $K_1 \approx 0.032 $ and $K_2 \approx 0.109$, and  for $\text{Bo} = 900$, and $K_1 \approx 0.017$ and $K_2 \approx 0.104$.  }
\label{fig:trans_torque_polyfit}
\end{figure}

\begin{figure}
    \centering
\includegraphics[scale=0.6]{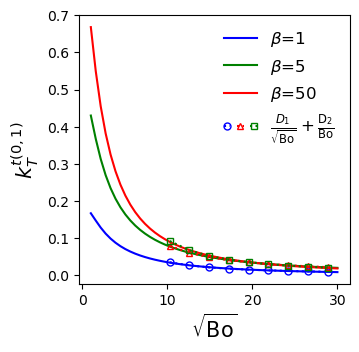}
    \caption{
    The torque coefficient $k_T^{(0,1)}$ in comparison with its large-Bo asymptotic approximation $D_1/\sqrt{\text{Bo}} + D_2/\text{Bo}$ plotted as a function of $\sqrt{\text{Bo}}$ with $\beta = 1, 5,$ and $50,$ where $D_1 \approx 1.3$ and $D_2 \approx 0.24$ for $\beta = 1$,  $D_1 \approx 2.8$ and $D_2 \approx 0.54$ for $\beta = 5$,  and  $D_1 \approx 5.56$ and $D_2 \approx 0.41$ for $\beta = 50$.}
    \label{fig:trans_torque_q_asym}
\end{figure}

The second problem is the asymptotic problem associated with the translational flow field $\mathbf{u}$ and stress $\bm{\sigma}$ up to orders Ca and $\delta$ [see equations \eqref{eqn:trans:asym_exp_for_vel} and \eqref{eqn:trans:asym_exp_for_stress}]. 
The rotational flow field $\mathbf{u}^{r(0,0)}$ and the truncated asymptotic series of the translational flow field $\mathbf{u}$ are both Stokes flows defined in the same geometry and therefore related by the Lorentz reciprocal theorem, 
\begin{align}
    \iint\limits_{\Sigma_P^{(0,0)} \cup \Sigma_I^{(0,0)} \cup \Sigma_\infty} \bm{\sigma}^{r(0,0)} \cdot \mathbf{n} \cdot \mathbf{u} \text{ d}\Sigma =\iint\limits_{\Sigma_P^{(0,0)} \cup \Sigma_I^{(0,0)} \cup  \Sigma_\infty} \bm{\sigma}^{} \cdot \mathbf{n} \cdot \mathbf{u}^{r(0,0)} \text{ d}\Sigma. \label{eqn:rot:reciprocal_theorem_relation}
\end{align} 
 It is worth noting that the rotational stress $\bm{\sigma}^{r(0,0)}$ tangent to  the particle surface exhibits a logarithmic singularity at the TCL \cite{oneill1986}. Since the integral of a logarithmically singular function converges in the classical Riemann sense, the reciprocal equation \eqref{eqn:rot:reciprocal_theorem_relation} remains valid.  Also, in the no-slip limit $\beta  \rightarrow \infty$, despite the logarithmic singularity in the rotational stress  $\bm{\sigma}^{r(0,0)}$  the translational flow velocity $\mathbf{u}^{(j,k)}$ tangent to the particle surface decays to zero sufficiently fast (as $\beta^{-1}$) so that the correction torques $T^{(j,k)}$ remain finite and valid for large $\beta$. 
By rearranging terms in Eq. \eqref{eqn:rot:reciprocal_theorem_relation} and following similar steps as in Sec. \ref{subsec:trans:drag_calculation} and Appendix \ref{app:reciprocal_theorem}, we can rewrite the integral \textcircled{\small 3} as integrals over the flat interface $\Sigma_I^{(0,0)}$. This allows us to derive explicit analytical expressions for the correction torques $T^{(j,k)}$ (see Appendix \ref{app:reciprocal_theorem:torque}), i.e., 
{
\begin{align}
\begin{split}
T^{(j,k)}= \iint_{\Sigma_P^{(0,0)}} \bm{\sigma}^{(j,k)} \cdot \hat{\mathbf{e}}_\rho \cdot \mathbf{V}^r \text{ d}\Sigma  & = \iint_{\Sigma_I^{(0,0)}} \bm{\sigma}^{r(0,0)}\cdot \hat{\mathbf{e}}_z\cdot \mathbf{u}^{(j,k)} \mbox{ d}\Sigma \\ & -  \iint_{\Sigma_I^{(0,0)}} \bm{\sigma}^{(j,k)}\cdot \hat{\mathbf{e}}_z\cdot \mathbf{u}^{r(0,0)} \mbox{ d}\Sigma.
\label{eqn:trans:reciprocal_eqn2}
 \end{split}
\end{align}}
In Eq. \eqref{eqn:trans:reciprocal_eqn2}, the rotational flow field $\mathbf{u}^{r(0,0)}$ and stress $\bm{\sigma}^{r(0,0)}$ are given by  Eqs. \eqref{eqn:app:order1_sol1} - \eqref{eqn:app:order1_sol4} from Ref. \cite{oneill1986}, and the $\mathcal{O}(\text{Ca})$ and $\mathcal{O}(\delta)$ translational flow field $\mathbf{u}^{(j,k)}$ and stress $\bm{\sigma}^{(j,k)}$ can be obtained from the interface conditions \eqref{eqn:trans:reciprocal:interface_condition1} and \eqref{eqn:trans:reciprocal:interface_condition2}. The $\mathcal{O}(\text{Ca})$ torque $T^{(1,0)}$ is integrated to be zero due to the similar antisymmetry arguments, and the $\mathcal{O}(\delta)$ torque $T^{(0,1)}$ is  integrated to be 
{ 
\begin{align}
\begin{split}
     T^{(0,1)}
     = &  \iint_{\Sigma_I^{(0,0)}} \sigma_{zz}^{r(0,0)} \left[\left(u_r^{(0,0)}-\cos\varphi\right) \frac{\text{d} h^{(0,1)}}{\text{d} r} - \frac{\partial u_z^{(0,0)}}{\partial z} h^{(0,1)} \right] \mbox{ d}\Sigma \\
     & -\iint_{\Sigma_I^{(0,0)}}   u_r^{r(0,0)}\left[-\frac{\partial \sigma_{zr}^{(0,0)}}{\partial z} h^{(0,1)} + (\sigma_{rr}^{(0,0)}-\sigma_{zz}^{(0,0)} ) \frac{\text{d} h^{(0,1)}}{\text{d} r}\right]  \\ &  + u_\varphi^{r(0,0)}\left[-\frac{\partial \sigma_{z\varphi}^{(0,0)}}{\partial z} h^{(0,1)} + \sigma_{\varphi r}^{(0,0)} \frac{\text{d} h^{(0,1)}}{\text{d} r} \right]  \mbox{ d}\Sigma.
 \end{split}\label{eqn:trans:torque_T^t(0,1)_formula}
\end{align}
} 
Here we rewrite $T^{(0,1)}$ as 
\begin{align}
    T^{(0,1)} = \Tilde{\theta}_s k^{(0,1)}_T (\sqrt{\text{Bo}},\beta), \label{eqn:trans:torque_coeff_def}
\end{align}
{
where $k^{(0,1)}_T$ is the $\mathcal{O}(\delta)$ resistive torque coefficient due to the particle translation. 
In Fig. \ref{fig:trans_torque_polyfit}, we plot the torque coefficient $k_T^{(0,1)}$ as a function of $\beta$ and varying values of Bond number. For a perfect slip on the particle surface ($\beta = 0$), the torque exerted by the translational flow vanishes and the torque coefficient $k_T^{(0,1)} = 0$. In Fig. \ref{fig:trans_torque_polyfit}(a), we observe that for $\text{Bo} = 1, 5,$ and $50,$ the torque coefficient $k_T^{(0,1)}$ monotonically increases and converges to a finite value as $\beta$ increases. For $\beta \gg 1,$ the asymptotic behavior of $k_T^{(0,1)}$ can be expressed as 
\begin{align}
    k_T^{(0,1)} \sim K_1 + \frac{K_2}{\beta},
\end{align}
where $K_1$ and $K_2$ can be determined through curve-fitting for a fixed value of Bo. We obtain  $K_1 \approx 0.71$ and $K_2 \approx -2.23$ for $\text{Bo} = 1$, $K_1 \approx 0.47$ and $K_2 \approx -1.21$ for $\text{Bo} = 5$, $K_1 \approx 0.15$ and $K_2 \approx -0.14$ for $\text{Bo} = 50$, and $K_1 \approx 0.1$ and $K_2 \approx -0.01$ for $\text{Bo} = 96.5$.  In the limit of $\beta \rightarrow \infty,$ $k_T^{(0,1)}$ converges to the $\mathcal{O}(\delta)$ translational torque with no-slip boundary conditions, which are approximated by the values of $K_1$. As we increase Bo, the dependence of $k_T^{(0,1)}$ on the slip coefficient $\beta$ becomes nonmonotonic. In Fig. \ref{fig:trans_torque_polyfit}(b), we observe that for $\text{Bo} = 200, 400,$ and $900$, as $\beta$ increases, $k_T^{(0,1)}$ initially increases, then decreases and  converge  to a finite value. This convergence for $\beta \gg 1$ also behaves asymptotically like $ k_T^{(0,1)} \sim K_1 + K_2/\beta,$ where $K_1$ and $K_2$ can be determined via curve-fitting. We find that $K_1 \approx 0.056$ and $K_2 \approx 0.081$ for $\text{Bo} = 200$, $K_1 \approx 0.032 $ and $K_2 \approx 0.109$ for $\text{Bo} = 400$, and $K_1 \approx 0.017$ and $K_2 \approx 0.104$ for $\text{Bo} = 900$. Here the values of no-slip torque coefficient $k_T^{(0,1)} (\sqrt{\text{Bo}},\beta \rightarrow \infty)$ are approximated by $K_1$. Note that the $\mathcal{O}(\delta)$ torque coefficient's nonmonotonic dependence on $\beta$ results from the coupling of the interfacial deformation and the translational and rotational flow fields. For small Bond numbers, the nonmonotonic torque contribution is dominated by the monotonic torque contribution. As the Bond number increases, the monotonic contribution decays more rapidly than the nonmonotonic contribution, resulting in the emergent nonmonotonic behavior. The critical Bond number at which the nonmonotonic $\beta$-dependence appears is numerically determined to be $\text{Bo} \approx 96.5$ (see Fig. \ref{fig:trans_torque_polyfit}). 

In Fig. \ref{fig:trans_torque_q_asym}, we plot the torque coefficient $k_T^{(0,1)}$ as a function of $\sqrt{\text{Bo}}$ for $\beta = 1,  5$ and 50. The torque coefficient $k_T^{(0,1)}$ exhibits the same qualitative dependence on the Bond number as that of the drag coefficient $k_D^{(0,1)}$. As the Bond number increases, an increase in the liquid density flattens the interface around the particle and, as a result, reduces the correction torque due to the static deformation. In the limit of large Bond number, the asymptotic behavior of $k_T^{(0,1)}$ can be expressed as 
\begin{align}
    k_T^{(0,1)} \sim \frac{D_1}{\sqrt{\text{Bo}}} + \frac{D_2}{\text{Bo}}, \label{eqn:trans:torque_coeff_large_Bo_asym}
\end{align}
where the values of $D_1$ and $D_2$ are determined through curve-fitting for a fixed value of $\beta.$ We obtain $D_1 \approx 1.3$ and $D_2 \approx 0.24$ for $\beta = 1$,  $D_1 \approx 2.8$ and $D_2 \approx 0.54$ for $\beta = 5$,  and  $D_1 \approx 5.56$ and $D_2 \approx 0.41$ for $\beta = 50$. The large-Bo asymptotic approximations \eqref{eqn:trans:torque_coeff_large_Bo_asym} is plotted in Fig. \ref{fig:trans_drag_q_asym} along with  torque coefficient $k_T^{(0,1)}$. In the limit $\text{Bo} \rightarrow 0,$ the $k_T^{(0,1)}$ becomes invalid as the static deformation fails to satisfy the far-field condition \eqref{eqn:far_field_condition_static_def}. 
}

\section{Pair interaction of particles}\label{sec:capillary_interactions}

In Ref \cite{danov2010}, Danov and Kralchevsky proposed a general theoretical approach for calculating the lateral capillary forces between two floating particles at a liquid interface by integration in the midplane, where the superposition approximation of the interfacial deformation can be obtained.  In our earlier work \cite{zhou2022}, we calculated the vertical capillary forces between two spherical particles at a deformable liquid interface undergoing a uniform background flow. The capillary forces exerted on the particles by the static and flow-induced deformations were numerically computed by directly integrating the the capillary stress along the TCLs.  D\"orr and Hardt  \cite{dorr2015} studied the pair interactions between two driven particles translating along a liquid interface with pinned TCLs. The interfacial deformation around the two particles is constructed from the single-particle deformation and the capillary interaction forces are defined and calculated by integrating the capillary stress along a path that runs through the interface halfway between the particles and at infinite distance from the particles, see Fig. \ref{fig:sketch-two-particle-EFMN}. 
With the analytical solutions of the interfacial deformations around a translating and rotating particle available to us here, we parallel D\"orr and Hardt's analysis to calculate the capillary interaction forces (defined in Ref. \cite{dorr2015}) and torques between two translating and rotating particles and give an analytical estimate of the interaction force for large separations. 

\subsection{Capillary interaction between two translating particles}\label{subsec:trans:capillary_interactions}
\begin{figure}
    \centering
    \includegraphics[width=8cm]{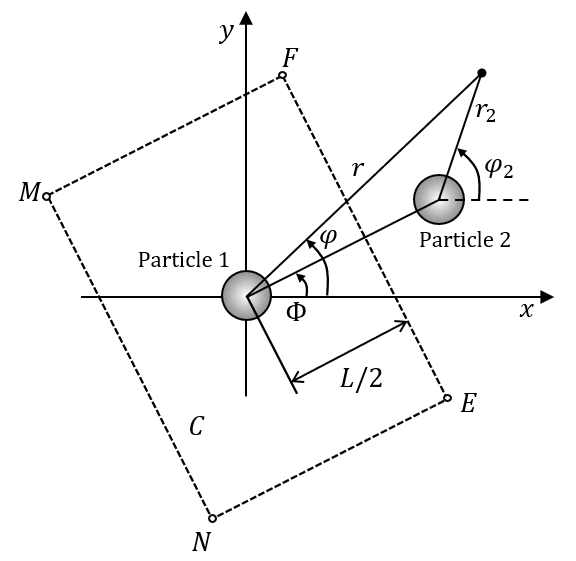}
    \caption{A sketch of two translating and rotating surface-trapped particles with separation distance $L$. The rectangle $EFMN$ lines in the plane of $z=0$.  The interaction forces and torques are evaluated along the integration path along the fluid interface $C = C_{EF} \cup C_{FM} \cup C_{MN} \cup C_{EN}$.  The path $C$ is found by projecting the rectangle $EFMN$ onto the deformed interface.}
    \label{fig:sketch-two-particle-EFMN}
\end{figure}

First consider the problem of two spherical particles translating along a gas-liquid interface with velocity $\hat{\mathbf{e}}_x$. This geometry is sketched in Fig. \ref{fig:sketch-two-particle-EFMN} where the two particles are shown resting on the interface at a separation distance $L$. Also illustrated is the rectangle $EFMN$ which lies on the undeformed surface $z=0$, encloses Particle 1, and whose side  $EF$ lies midway between the two particles.

As a notational convenience, we use the superscript ``$t$" to denote variables in the translational problem and ``$r$"  the rotational problem. 
Assuming a large separation distance between the two particles, the leading-order far-field interfacial deformation comes from the flow-induced deformation $h^{t(0,1)}$ and is given by (see calculation details in Appendix \ref{App:far-field_approx_of_flow_induced_defs}),
\begin{align}
   & \text{Ca} h^{t(1,0)} \sim \frac{\text{Ca} 3\beta}{2(3+\beta)} \frac{1}{\text{Bo} r^4} \cos\varphi,
   \label{eq:63}
\end{align}
which is valid for $\sqrt{\text{Bo}} r \gg 1$ and for a single particle.
Let  $\Phi$ denote the orientation angle, which is the angle between the line-of-center of the two particles and the $x$-axis. To describe the two-sphere system, we introduce the  cylindrical coordinates $(r_i,\varphi_i,z)$, where the origin is centered at the center of mass of particle $i,$ $i = 1,2$ (see Fig. \ref{fig:sketch-two-particle-EFMN}). Then, the flow-induced deformations from particle $i$ for $\sqrt{\text{Bo}}r_i \gg 1$ can be described by \eqref{eq:63} where $r$ is replaced by $r_i$.

We now combine the two cylindrical coordinates $(r_i,\varphi_i,z)$, $i=1,2$, by applying the following coordinate transformation:
\begin{align}
    r_1 = r, \quad \varphi_1 = \varphi,\quad z = z, 
    \label{eq:64}
\end{align}
and \begin{align}
    r_2 = \sqrt{r^2 + L^2 - 2 L r \cos(\varphi - \Phi)}, \quad \varphi_2 = \arccos{\frac{r\cos\varphi - L \cos\Phi}{r_2}}, \quad z_2 = z.
    \label{eq:65}
\end{align}
Then, the LSA of the two-particle deformation is  given by 
\begin{align}
    \mbox{Ca}h_{LSA}^{t(1,0)} = \mbox{Ca} \left(  h_1^{t(1,0)} + h_2^{t(1,0)}  \right),
    \label{eq:66}
\end{align}
where 
\begin{align}
   & h_1^{t(1,0)} \approx  \frac{3\beta}{2(3+\beta)} \frac{1}{\text{Bo} r^4} \cos\varphi, \quad h_2^{t(1,0)} \approx   \frac{3\beta}{2 (3+\beta)\text{Bo}} \frac{r \cos\varphi - L \cos\Phi}{ (r^2 + L^2 + 2 L r \cos(\varphi - \Phi))^{5/2}}.
   \label{eq:67}
\end{align}
Let $\tilde{\mathbf{F}}_{int}^t$ denote the dimensional interaction force acting on one of the particles. By paralleling the analysis in Ref. \cite{dorr2015}, we can rewrite  $\mathbf{F}_{int}^t$, the nondimensionalized interaction force due to the flow-induced deformation $\text{Ca} h^{t(0,1)}$, as 
\begin{align}
  \mathbf{F}_{int}^t   = \frac{\tilde{\mathbf{F}}^t_{int}}{\mu U a} = \frac{\gamma a}{\mu U a} \oint_C  \frac{\partial \mathbf{r}_C}{\partial s} \times \hat{\mathbf{n}}_{LSA}  \mbox{ d}s = \frac{1}{\text{Ca}} \oint_C  \frac{\partial \mathbf{r}_C}{\partial s} \times \hat{\mathbf{n}}_{LSA} \mbox{ d}s,\label{eqn:trans:cap_force_formula_1}
\end{align}
where $\text{Ca} = \mu U/\gamma$, $\mathbf{r}_C$ is the parametrization vector for a point on the integration path $C$ (see Figure \ref{fig:sketch-two-particle-EFMN}), and  $\hat{\mathbf{n}}_{LSA}$ is the unit normal vector to the fluid interface truncated after the $\mathcal{O}(\mbox{Ca})$ term, i.e.,
\begin{align}
    \hat{\mathbf{n}}_{LSA} = \hat{\mathbf{e}}_z - \mbox{Ca}\left(\frac{\partial h_{LSA}^{t(1,0)}}{\partial r} \hat{\mathbf{e}}_r + \frac{1}{r}\frac{\partial  h_{LSA}^{t(1,0)}}{\partial \varphi}\right).  
\end{align}

Here, the integration path $C = C_{EF} \cup C_{FM} \cup C_{MN} \cup C_{EN}$ is found by projecting the rectangle $EFMN$ onto the deformed interface approximated by $h^{t(1,0)}_{LSA}$. Note that $C_{EF}$ is located halfway between the two particles, and $C_{FM}, C_{MN},$ and $C_{EN}$ will be sent to infinity. 

Similarly, the nondimensionalized interaction torque $\mathbf{T}_{int}^t$ can be expressed as 
\begin{align}
   \mathbf{T}_{int}^t   = \frac{\tilde{\mathbf{T}}^t_{int}}{\mu U a^2} = \frac{\gamma a^2}{\mu U a^2} \oint_C  \mathbf{r}_c\times\left(\frac{\partial \mathbf{r}_c}{\partial s} \times \hat{\mathbf{n}}_{LSA} \right) \mbox{ d}s = \frac{1}{\text{Ca}} \oint_C\mathbf{r}_c\times\left(\frac{\partial \mathbf{r}_c}{\partial s} \times \hat{\mathbf{n}}_{LSA} \right)  \mbox{ d}s, \label{eqn:trans:cap_torque_formula_1}
\end{align}
where $\tilde{\mathbf{T}}_C^t$ denotes the dimensional interaction torque. 

In Eq. \eqref{eqn:trans:cap_force_formula_1}, the contributions from the integrals along paths $C_{FM}$ and $C_{EN}$ at infinite distance from the particles vanish because the deformation $h^{t(1,0)}_{LSA}$ decays to zero at infinity. To evaluate the sum of the integrals along $C_{EF}$ and $C_{MN}$, we first consider the following parametrization along $C_{EF}$,  
\begin{align}
    \mathbf{r}_C = \frac{L}{2\cos(\varphi-\Phi)} \hat{\mathbf{e}}_r + \mbox{Ca} h^{t(1,0)}_{LSA} \vert_{r = L/2\cos(\varphi-\Phi)}\hat{\mathbf{e}}_z.
\end{align}
Then, we account for the contribution from $C_{MN}$ by subtracting the vector of constant orientation 
\begin{align}
    \frac{\partial}{\partial \varphi} \left(\frac{L}{2\cos(\varphi - \Phi)}\hat{\mathbf{e}}_r \right) \times \hat{\mathbf{e}}_z = \frac{L}{2\cos(\varphi-\Phi)}\hat{\mathbf{e}}_r - \frac{L\sin(\varphi-\Phi)}{2\cos^2(\varphi-\Phi)} \hat{\mathbf{e}}_\varphi
\end{align}
from the integrand along $C_{EF}$ \cite{dorr2015}. 
Finally, the interaction force is calculated to be
\begin{align}
\begin{split}
   \mathbf{F}_{int}^t   =  &  \frac{1}{\text{Ca}}
    \int_{C_{EF}} \frac{\partial \mathbf{r}_{C_{EF}}}{\partial s} \times \hat{\mathbf{n}}_{LSA}   -  \frac{\partial}{\partial \varphi} \left(\frac{L}{2\cos(\varphi - \Phi)}\hat{\mathbf{e}}_r \right) \times \hat{\mathbf{e}}_z \mbox{ d}s \\
     = &\frac{1}{\text{Ca}} \int_{\Phi-\pi/2}^{\Phi + \pi/2} \frac{\mbox{d} \mathbf{r}_{C_{EF}}}{\mbox{d} \varphi} \times \left.\left( \hat{\mathbf{e}}_z - \mbox{Ca}\left(\frac{\partial h_{LSA}^{(1,0)}}{\partial r} \hat{\mathbf{e}}_r + \frac{1}{r}\frac{\partial  h_{LSA}^{(1,0)}}{\partial \varphi}\right) \right)\right\vert_{r = L/2\cos(\varphi - \Phi)} \\ - & \frac{\partial}{\partial \varphi} \left(\frac{L}{2\cos(\varphi - \Phi)}\hat{\mathbf{e}}_r \right) \times \hat{\mathbf{e}}_z \mbox{ d}\varphi  \\
    = & -\frac{192  \beta}{L^4 \text{Bo} (3+\beta)} \cos\Phi \hat{\mathbf{e}}_z+ \frac{3465\mbox{Ca} \pi \beta^2 \cos\Phi\sin^2\Phi}{(3+\beta)^2 L^9 \text{Bo}^2} \hat{\mathbf{e}}_x  \\ & - \frac{315\mbox{Ca} \pi \beta^2 (\sin\Phi+11\sin(3\Phi))}{4(3+\beta)^2 L^9 \text{Bo}^2} \hat{\mathbf{e}}_y. 
    \end{split}\label{eqn:trans:cap_force_formula_2}
\end{align}

The calculation of the interaction torque \eqref{eqn:trans:cap_torque_formula_1} is analogous to that of the interaction force, which gives 
\begin{align}
\begin{split}
    \mathbf{T}^t_{int} = & \frac{1}{\text{Ca}}  \int_{C_{EF}} \mathbf{r}_{C_{EF}} \times\left(\frac{\mathbf{r}_{C_{EF}}}{\partial s} \times \hat{\mathbf{n}}_{LSA}  \right) \\ & - \frac{L}{2\cos(\varphi-\Phi)} \hat{\mathbf{e}}_r \times\left(\frac{\partial}{\partial \varphi} \left(\frac{L}{2\cos(\varphi - \Phi)}\hat{\mathbf{e}}_r \right) \times \hat{\mathbf{e}}_z\right)\mbox{ d}s\\
    = & -\frac{5355\mbox{Ca} \pi \beta^2 \sin(2\Phi) }{8(3+\beta)^2 L^8 \text{Bo}^2}\hat{\mathbf{e}}_z -\frac{48  \beta \sin (2\Phi)}{(3+\beta) L^3 \text{Bo}} \hat{\mathbf{e}}_x + \frac{96  \beta \cos^2\Phi}{(3+\beta) L^3 \text{Bo}}\hat{\mathbf{e}}_y. 
    \end{split}\label{eqn:trans:cap_torque_formula_2}
\end{align}

\subsection{Capillary interaction between two rotating particles}
Second, we consider the problem of two rotating particles at a gas-liquid interface with angular velocity $\hat{\mathbf{e}}_y$. 
The $\mathcal{O}(\text{Ca})$ deformation induced by the particle's rotational motion, Ca$h^{r(1,0)}$, can be obtained by paralleling the calculation in Sec. \ref{sec:interfacial_deformations} and is given in Appendix \ref{App:far-field_approx_of_flow_induced_defs}.
Adopting similar notations, the LSA of the flow-induced deformation $\text{Ca} h^{r(0,1)}$ is given by (see Appendix \ref{App:rot-flow-induced-deformation} for derivation details)
\begin{align}
    \mbox{Ca}h_{LSA}^{r(1,0)} = \frac{\text{Ca} \mathcal{D}(\beta)}{\text{Bo}}\left(   \frac{1}{ r^4} +   \frac{r \cos\varphi - L \cos\Phi}{ (r^2 + L^2 + 2 L r \cos(\varphi - \Phi))^{5/2}} \right)\cos\varphi,
\end{align}
where 
\begin{align}
    \mathcal{D}(\beta) = -\frac{3\beta (7\beta^2 + 83 \beta + 210)}{8(\beta+3)(\beta+4)(\beta+7)}.
\end{align}
The interaction force and torque due to the deformation induced by particle rotation can be written as 
\begin{align}
   & \mathbf{F}_{int}^r = \frac{\tilde{\mathbf{F}}^r_{int}}{\mu U a^2} = \frac{\gamma a^2}{\mu U a^2} \oint_C  \frac{\partial \mathbf{r}_c}{\partial s} \times \hat{\mathbf{n}}_{LSA}  \mbox{ d}s = \frac{1}{\text{Ca}} \oint_C  \frac{\partial \mathbf{r}_c}{\partial s} \times \hat{\mathbf{n}}_{LSA}  \mbox{ d}s, \label{eqn:rot:cap_force_formula_1} \\
   &  \mathbf{T}_{int}^r = \frac{\tilde{\mathbf{T}}^r_{int}}{\mu U a^3} = \frac{\gamma a^2}{\mu U a^3} \oint_C  \mathbf{r}_c\times\left(\frac{\partial \mathbf{r}_c}{\partial s} \times \hat{\mathbf{n}}_{LSA} \right) \mbox{ d}s = \frac{1}{\text{Ca}} \oint_C\mathbf{r}_c\times\left(\frac{\partial \mathbf{r}_c}{\partial s} \times \hat{\mathbf{n}}_{LSA} \right)  \mbox{ d}s. \label{eqn:rot:cap_torque_formula_1}
\end{align}
where $\text{Ca} = \mu a\Omega/\gamma$ and $\hat{\mathbf{n}}_{LSA} = \hat{\mathbf{e}}_z - \mbox{Ca}\left(\partial h_{LSA}^{r(1,0)}/\partial r \hat{\mathbf{e}}_r + 1/r\partial  h_{LSA}^{r(1,0)}/\partial \varphi\right).$
Paralleling the calculations of interaction force and torque in Sec. \ref{subsec:trans:capillary_interactions}, we obtain 
\begin{align}
\begin{split}
     \mathbf{F}_{int}^r =   & -\frac{128  \mathcal{D}(\beta)}{L^4 \text{Bo}} \cos\Phi \hat{\mathbf{e}}_z+ \frac{1540\mbox{Ca} \mathcal{D}(\beta)^2 \pi  \cos\Phi\sin^2\Phi}{ L^9 \text{Bo}^2} \hat{\mathbf{e}}_x  \\ & - \frac{35\mbox{Ca}  \mathcal{D}(\beta)^2 \pi (\sin\Phi+11\sin(3\Phi))}{L^9 \text{Bo}^2} \hat{\mathbf{e}}_y, \label{eqn:rot:cap_force_formula_2} 
     \end{split} \\
        \mathbf{T}_{int}^r = & -\frac{595\mbox{Ca}  \mathcal{D}(\beta)^2 \pi \sin(2\Phi) }{2 L^8\text{Bo}^2 }\hat{\mathbf{e}}_z -\frac{64  \mathcal{D}(\beta) \cos\Phi\sin\Phi}{L^3 \text{Bo}} \hat{\mathbf{e}}_x + \frac{64  \mathcal{D}(\beta) \cos^2\Phi}{ L^3 \text{Bo}}\hat{\mathbf{e}}_y.\label{eqn:rot:cap_torque_formula_2}
\end{align}

\subsection{Remarks about the Interaction Force}

Following the multipole analysis in Ref. \cite{dorr2015}, the interfacial deformations $h^{t,r(1,0)}$ due to the particle motions in Eq. \eqref{eqn:flow_induced_deformations_solutions} can be identified as dipolar in terms of the azimuthal angle $\varphi,$ and the vertical interaction forces due to the deformations, $\mathbf{F}_{int}^{t,r}$, in Eqs. \eqref{eqn:trans:cap_force_formula_2} and \eqref{eqn:rot:cap_force_formula_2} can also be labeled as dipolar interaction forces. Eqs. \eqref{eqn:trans:cap_force_formula_2} and \eqref{eqn:rot:cap_force_formula_2} show that the 
$\mathcal{O}(\text{Ca}^0)$ interaction forces act in the vertical ($\hat{\mathbf{e}}_z$) direction and decays as $L^{-4}$ in the limit of $L\rightarrow\infty,$ and the $\mathcal{O}(\text{Ca})$ interaction forces act in the $x$-y plane and decays as $L^{-9}$. In Ref. \cite{dorr2015}, D\"orr and Hardt assumed that the particles have pinned TCLs and are rotated until the torque balances are reached. They also assumed that the gravity effect is negligible ($\text{Bo}=0$). Compared to Eqs. \eqref{eqn:trans:cap_force_formula_2} and \eqref{eqn:rot:cap_force_formula_2}, the interaction force obtained  by D\"orr and Hardt  [Eq. (5.8) in Ref. \cite{dorr2015} ] exhibits a slower decay as $L\rightarrow \infty,$ where  the vertical interaction force $\mathbf{F}_{int} \cdot \hat{\mathbf{e}}_z$ decays as $L^{-2}$ and the lateral interaction force $\mathbf{F}_{int} \cdot \hat{\mathbf{e}}_{x,y}$ decays as $L^{-5}.$ The difference in decay rates is expected as the gravity force exerted by the liquid phase has a flattening effect on the interface and localizes the interfacial deformation.

In the case where $\text{Bo} = 0$, the far-field approximations of the flow-induced deformations \eqref{eqn:flow_induced_deformations_solutions} behave as  $h^{t,r(1,0)} \sim r^{-1},$ which is different from D\"orr and Hardt's deformation solution [Eq. (4.18) in Ref. \cite{dorr2015}], where the asymptotically dominant term in the far field decays as $r^{-2}$. This is because, under the assumptions made in Ref. \cite{dorr2015}, the particle is allowed to freely rotate until the torque balance is reached. The balancing of the hydrodynamic and capillary torques eliminates the $r^{-1}$-term. Consequently, in the absence of gravity effects, the lateral capillary interaction forces $\mathbf{F}_{int}^{t,r} \cdot \hat{\mathbf{e}}_{x,y}$ decays as $L^{-3}$ (see Appendix H for details).

In addition, we observe in Eqs. \eqref{eqn:trans:cap_force_formula_2} and \eqref{eqn:rot:cap_force_formula_2} that when the particles translate or rotate in the direction perpendicular to their line of centers ($\Phi = \pi/2$), the interaction forces are  attractive and act along the line of centers.  
The forces acting along the particles' line of centers are given by
\begin{align}
    &  \mathbf{F}^t_\parallel =  \mathbf{F}^t_{int}\cdot (\cos\Phi \hat{\mathbf{e}}_x + \sin\Phi\hat{\mathbf{e}}_y) = \frac{1575 \mbox{Ca} \pi \beta^2 (1-\cos(2\Phi))}{4 (3+\beta)^2 L^9 \text{Bo}^2}, \\
        &  \mathbf{F}^r_\parallel  = \mathbf{F}^r_{int}\cdot (\cos\Phi \hat{\mathbf{e}}_x + \sin\Phi\hat{\mathbf{e}}_y) = \frac{175 \mbox{Ca} \pi \mathcal{D}(\beta)^2 (1-\cos(2\Phi)}{   L^9 \text{Bo}^2}, 
\end{align}
where $\mathbf{F}^{t,r}_\parallel$ vanishes when the particles translate or rotate in the direction parallel to their line of centers.

It is also worth noting that in Ref. \cite{dorr2015}'s analysis, the hydrodynamic interaction between the moving particles is neglected due to the argument that the hydrodynamic forces acting on the two particles are equal. In Refs. \cite{Goldman1966, Stimson1926, zhou2022}, the hydrodynamic forces exerted on two translating particles in a homogeneous fluid or at a fluid-fluid interface are calculated and shown to be equal. However, it is necessary to point out that due to the presence of flow, the interaction forces $\mathbf{F}^{t,r}_{int}$  and torques $\mathbf{T}^{t,r}_{int}$ defined in this section do not represent the capillary forces obtained from integration along the TCLs. Indeed, in the case of two floating particles at an interface, the capillary force can be expressed alternatively via integration in the midplane halfway between the two particles by taking advantage of the force balance between interfacial tension and hydrostatic pressure (see Eq. (2.5) in Ref. \cite{danov2010}).  This is not true in the presence of flow as the contribution from the two-particle hydrodynamic stress also must be considered. Further analytical or numerical calculation is required to obtain the exact capillary forces and torques exerted on the two particles.

\subsection{Total vertical interaction force between two translating particles}\label{app:total_int_force}
Here, we apply Ref. \cite{danov2010}'s analytical method to the problem of two translating particles at a liquid interface, which allows us to obtain an asymptotic approximation for the total interaction forces acting on the translating particles. 
We define the total interaction force acting on one of the two particles as the total force acting on one of the two particles subtracted by the total force acting on an isolated particle, i.e., 
\begin{align}
    \mathbf{F}_{tot} = \mathbf{F}_{2p} - \mathbf{F}_{1p} \label{eqn:total_int_force:tot_int_force1}.
\end{align}
Here, $\mathbf{F}_{1p}$ and $\mathbf{F}_{2p}$ denote the respective nondimensionalized single-particle and two-particles total forces and are given by 
\begin{align}
    \mathbf{F}_{1p} = &  \frac{1}{\text{Ca}}\int_{\Sigma_{\text{TCL},1p}} \tilde{\mathbf{n}}_{C,1p} \text{ d}s +  \iint_{\Sigma_{P,1p}}\bm{\sigma}_{1p}\cdot(-\hat{\mathbf{e}}_\rho) \mbox{ d}\Sigma - \frac{\text{Bo}}{\text{Ca}} \iint_{\Sigma_{P,1p}} \mathbf{P}\cdot(-\hat{\mathbf{e}}_\rho) \mbox{ d}\Sigma, \label{eqn:total_int_force:F1p_definition}\\
        \mathbf{F}_{2p} = &  \frac{1}{\text{Ca}}\int_{\Sigma_{\text{TCL},2p}} \tilde{\mathbf{n}}_{C,2p} \text{ d}s +  \iint_{\Sigma_{P,2p}}\bm{\sigma}_{2p}\cdot(-\hat{\mathbf{e}}_\rho) \mbox{ d}\Sigma - \frac{\text{Bo}}{\text{Ca}} \iint_{\Sigma_{P,2p}} \mathbf{P}\cdot(-\hat{\mathbf{e}}_\rho) \mbox{ d}\Sigma,\label{eqn:total_int_force:F2p_definition}
    \end{align}
where $\tilde{\mathbf{n}}_{C,1p}$ and $\tilde{\mathbf{n}}_{C,2p}$ denote the capillary unit vector normal to the TCLs and lie in the interface, $\bm{\sigma}_{1p}$ and $\bm{\sigma}_{2p}$ are the dimensionless single-particle and two-particle viscous stress tensors, and $\mathbf{P} = z \mathbf{I}$ is the dimensionless hydrostatic pressure tensor. The leading-order approximation of Eq. \eqref{eqn:total_int_force:F1p_definition} in the $\hat{\mathbf{e}}_z$-component is given by the single-particle capillary force calculation and direct integral evaluations with the zeroth-order viscous stress  and hydrostatic pressure tensors (see \cite{zhou2022,oneill1986,phdApril}), i.e.,
\begin{align}
    \hat{\mathbf{e}}_z\cdot\mathbf{F}_{1p} \approx \frac{\delta}{\text{Ca}} 2\pi \tilde{\theta}_s \sqrt{\text{Bo}} C_0 K_1(\sqrt{\text{Bo}}) - \frac{\text{Bo}}{\text{Ca}} \frac{2\pi}{3},\label{eqn:total_int_force:F1p_solution}
\end{align}
where $C_0$ is given in Eq. \eqref{eqn:static_deformation}.

Moving onto the calculation of the two-particle total force $\mathbf{F}_{2p}$, we first note that at the liquid interface $z = h_{2p}(r,\varphi)$, the viscous stress is balanced by the capillary stress and the hydrostatic pressure, i.e.,
\begin{align}
    \frac{1}{\text{Ca}} (\nabla\cdot\hat{\mathbf{n}}_{2p})\hat{\mathbf{n}}_{2p} + \frac{\text{Bo}}{\text{Ca}} \mathbf{P} \cdot \hat{\mathbf{n}}_{2p}=  \bm{\sigma}_{2p} \cdot \hat{\mathbf{n}}_{2p},\label{eqn:total_int_force:stress_balance1}
\end{align}
where $\hat{\mathbf{n}}$ denotes the unit normal to the liquid interface $z = h_{2p}(r,\varphi)$
 pointing towards the liquid. 
Here, we adopt similar notations to those used in Ref. \cite{danov2010} and let $\Sigma_{EFMN}$ denote the deformed interface bounded by the curve $C$ and particle, see  Fig. \ref{fig:sketch-two-particle-EFMN}.
Rearranging and integrating \eqref{eqn:total_int_force:stress_balance1} over the surface  $\Sigma_{EFMN}$, we obtain 
\begin{align}
       \frac{1}{\text{Ca}} \iint_{\Sigma_{EFMN}}(\nabla\cdot\hat{\mathbf{n}}_{2p})\hat{\mathbf{n}}_{2p}\mbox{ d}\Sigma = \iint_{\Sigma_{EFMN}} \left(\bm{\sigma} - \frac{\text{Bo}}{\text{Ca}}\mathbf{P} \right) \cdot \hat{\mathbf{n}}_{2p} \mbox{ d}\Sigma. 
 \label{eqn:total_int_force:stress_balance2}
\end{align}
By applying the Stokes theorem to the left-hand side of Eq.  \eqref{eqn:total_int_force:stress_balance2}, we can rewrite Eq.   \eqref{eqn:total_int_force:stress_balance2} as 
\begin{align}
 \mathbf{F}_{C,2p} - \frac{1}{\text{Ca}} \int_C  \tilde{\mathbf{n}}_{C,2p}\mbox{ d}s = \iint_{\Sigma_{EFMN}} \bm{\sigma} \cdot \hat{\mathbf{n}}_{2p}\mbox{ d}\Sigma - \frac{\text{Bo}}{\text{Ca}} \iint_{\Sigma_{EFMN}}\mathbf{P}  \cdot \hat{\mathbf{n}}_{2p} \mbox{ d}\Sigma,\label{eqn:total_int_force:stress_balance3}
\end{align}
where $\mathbf{F}_{C,2p}$ is the two-particle capillary force defined in Eq. \eqref{eqn:total_int_force:F2p_definition}, i.e.,
\begin{align}
    \mathbf{F}_{C,2p}  = \frac{1}{\text{Ca}}\int_{\Sigma_{\text{TCL},2p}} \tilde{\mathbf{n}}_{C,2p} \text{ d}s.\label{eqn:total_int_force:capillary_force_definition1}
\end{align}
We consider the asymptotic expansion \eqref{eqn:asym_exp_f} for small Ca and $\delta$ and assume a large particle separation distance, i.e., $L \gg 1.$ Then, the LSA of leading-order interfacial deformation far away from the particles is given by Eqs. \eqref{eq:66}-\eqref{eq:67}. 
Similarly, the two-particle viscous stress $\bm{\sigma}_{2p}^{(0,0)}$ at order 1 can be constructed via LSA and is given  by 
\begin{align}
     \bm{\sigma}_{2p}^{(0,0)} \approx \bm{\sigma}_{LSA}^{(0,0)} = \bm{\sigma}^{(0,0)}_{1p}(r_1,\varphi_1,z) + \bm{\sigma}_{1p}^{(0,0)}(r_2,\varphi_2,z),
\end{align}
where $\bm{\sigma}^{(0,0)}$ is the zeroth-order stress around an isolated particle given in Appendix \ref{App:order1_solution}, and the spatial variables $(r_i,\varphi_i,z)$ are defined in Eqs.  \eqref{eq:64} and \eqref{eq:65} for $i = 1,2.$

Note that the second term on the left-hand side of Eq. \eqref{eqn:total_int_force:stress_balance3} is exactly the interaction force $\mathbf{F}_{int}^t$ defined and computed in section \ref{subsec:trans:capillary_interactions}. The zeroth-order asymptotic approximation of $\mathbf{F}_{int}^{t}$ in the $\hat{\mathbf{e}}_z$-component is given by
\begin{align}
    \frac{1}{\text{Ca}} \int_C \tilde{\mathbf{n}}_{2p} \mbox{ d}s \approx  \hat{\mathbf{e}}_z \cdot \mathbf{F}_{int}^{t} = \frac{192\beta  \cos\Phi}{L^4 \text{Bo} (3+\beta)} . \label{eqn:total_int_force:dorr_int_force_solution}
\end{align}
Equations \eqref{eqn:total_int_force:stress_balance3} and \eqref{eqn:total_int_force:dorr_int_force_solution} yield the leading-order asymptotic expression of the two-particle capillary force $\mathbf{F}_{C,2p}$ in the vertical component,
\begin{align}
    \hat{\mathbf{e}}_z \cdot \hat{\mathbf{F}}_{C,2p} \approx \frac{192 \beta \cos\Phi}{L^4 \text{Bo} (3+\beta)} + \iint_{\Sigma_{EFMN}^{(0,0)}}\bm{\sigma}_{2p}^{(0,0)}\cdot \hat{\mathbf{n}}_{2p} \text{ d}\Sigma - \frac{\text{Bo}}{\text{Ca}} \iint_{\Sigma_{EFMN}^{(0,0 )}}\left(\text{Ca}h^{(1,0)}_{2p} + \delta h^{(0,1)}_{2p} \right)   \text{ d}\Sigma,\label{eqn:total_int_force:capillary_force_definition2}
\end{align}
where  $\Sigma_{EFMN}^{(0,0)}$ denotes the flat fluid interface $z = 0$ enclosed by the rectangle $EFMN.$ 

To evaluate the second integral in Eq. \eqref{eqn:total_int_force:capillary_force_definition2}, we take advantage of the fact that the viscous stress $\bm{\sigma}_{2p}$ is divergence-free and obtain 
\begin{align}
    0 = &  \iiint_{V} \nabla\cdot \bm{\sigma}^{(0,0)}_{2p}\mbox{ d}V \\
    = & \left( \iint_{\Sigma_{EFMN}} + \iint_{\Sigma_{EF}  \cup \Sigma_{MN}}  + \iint_{\Sigma_{EN}  \cup \Sigma_{FM}}  +  \iint_{\Sigma_P} + \iint_{\Sigma_B} \right ) \bm{\sigma}_{2p}^{(0,0)}  \cdot \mathbf{n}\mbox{ d}\Sigma,\label{eqn:total_int_force:stress_divergence_free}
\end{align}
where $V$ is the right prism built on the rectangle $EFMN$ immersed in the liquid phase. It is enclosed by the particle surface $\Sigma_P$, the liquid interface $\Sigma_{EFMN}^{(0,0)}$, and the vertical planes $\Sigma_{EF}^{(0,0)}  \cup \Sigma_{MN}^{(0,0)}$ and $\Sigma_{EN}^{(0,0)} \cup \Sigma_{FM}^{(0,0)}$ that pass through the line segments  $EF$,  $MN$, $EN$ and $FM$, respectively, plus the bottom of the prism $\Sigma_B$, and are immersed in the liquid phase.  In terms of the coordinates, we could define $V$ as the region bounded by $-X_0 < x < L/2$, $-Y_0 < y < Y_0$, and $0 < z < Z_0$ where we account for the particle surface along $z=0$. 
In Eq. \eqref{eqn:total_int_force:stress_divergence_free},  $\mathbf{n}$ denotes the outward unit vector to the boundary of $V$. In order to evaluate the integrals in \eqref{eqn:total_int_force:stress_divergence_free} we will need to allow for $X_0$, $Y_0$, and $Z_0$ to tend to infinity.  In the work presented below, these improper integrals are evaluate by first letting $Y_0 \rightarrow \infty$, then $X_0 \rightarrow \infty$, and finally $Z_0 \rightarrow \infty$.  When this is done, we find no contribution from $\Sigma_B$, and $\Sigma_{FM}^{(0,0)}\cup\Sigma_{EN}^{(0,0)}$.

Note that the vertical two-particle viscous force defined in \eqref{eqn:total_int_force:F2p_definition} can be rewritten as  
\begin{align}
    \hat{\mathbf{e}}_z \cdot \mathbf{F}_{V,2p} = &  \hat{\mathbf{e}}_z \cdot \iint_{\Sigma_{P,2p}} \bm{\sigma}_{2p}\cdot\hat{\mathbf{e}}_\rho\mbox{ d}\Sigma \\
    \approx & \hat{\mathbf{e}}_z \cdot \iint_{\Sigma_{P}^{(0,0)}} \bm{\sigma}_{2p}^{(0,0)}\cdot(-\hat{\mathbf{e}}_\rho)\mbox{ d}\Sigma \\
     = & -\hat{\mathbf{e}}_z \cdot \iint_{\Sigma_{EF}^{(0,0)}\cup\Sigma_{MN}^{(0,0)}}  \bm{\sigma}_{LSA,2p}^{(0,0)} \cdot\mathbf{n}\text{ d}\Sigma - \hat{\mathbf{e}}_z \cdot \iint_{\Sigma_{EFMN}^{(0,0)}}  \bm{\sigma}^{(0,0)}_{2p}\cdot \hat{\mathbf{n}}_{2p} \text{ d}\Sigma\\
    = & -\frac{8\beta \cos\Phi}{L^2(3+\beta)}  -  \hat{\mathbf{e}}_z \cdot \iint_{\Sigma_{EFMN}^{(0,0)}}  \bm{\sigma}^{(0,0)}_{2p}\cdot \hat{\mathbf{n}}_{2p} \text{ d}\Sigma,\label{eqn:total_int_force:visocus_force1}
\end{align}
where we have sent all sides to infinity.
Summing Eq. \eqref{eqn:total_int_force:capillary_force_definition2} and \eqref{eqn:total_int_force:visocus_force1} gives 
\begin{align}
    \hat{\mathbf{e}}_z \cdot \left(\mathbf{F}_{C,2p} + \mathbf{F}_{V,2p}\right) \approx \frac{192\beta \cos\Phi}{L^4 \text{Bo} (3+\beta) } - \frac{8\beta \cos\Phi}{L^4(3+\beta)} -  \frac{\text{Bo}}{\text{Ca}} \iint_{\Sigma_{EFMN}^{(0,0 )}}\left(\text{Ca}h^{(1,0)}_{2p} + \delta h^{(0,1)}_{2p} \right)   \text{ d}\Sigma.\label{eqn:total_int_force:sum_capillary_viscous_forces}
\end{align}
Note that at leading order, the two-particle hydrostatic pressure force $\mathbf{F}_{P,2p}$ in the vertical direction equals to the vertical single-particle hydrostatic pressure force, i.e.,
\begin{align}
   \hat{\mathbf{e}}_z\cdot  \mathbf{F}_{P,2p} = \frac{\text{Bo}}{\text{Ca}} \hat{\mathbf{e}}_z \cdot \iint_{\Sigma_{P,2p}} \mathbf{P}\cdot \hat{\mathbf{e}}_\rho \text{ d}\Sigma = & \int_0^{2\pi} \int_{\text{Ca}h^{(1,0)}_{2p} + \delta h^{(1,0)}_{2p} } ^1 z \sqrt{1-z^2} \text{ d}z \text{ d}\varphi \\
   \approx & \frac{\text{Bo}}{\text{Ca}}\frac{2\pi}{3}.
\end{align}
This implies that the hydrostatic pressure does not contribute to the total interaction force $\mathbf{F}_{tot}$ defined in Eq. \eqref{eqn:total_int_force:tot_int_force1}, and that 
$\hat{\mathbf{e}}_z \cdot \mathbf{F}_{tot}  = 
\hat{\mathbf{e}}_z\cdot   \left( \mathbf{F}_{C,2p} +\mathbf{F}_{V,2p}\right).$
Then, according to Eq. \eqref{eqn:total_int_force:sum_capillary_viscous_forces}, we only have the surface integral over $\Sigma_{EFMN}^{(0,0)}$ (last term in Eq. \eqref{eqn:total_int_force:sum_capillary_viscous_forces}) left to calculate. Since the LSA approximation of the two-particle deformation  is only accurate far away from the particles, here we additionally assume that $1\ll \text{Bo}\ll \text{Ca}^{-1}.$ Then, the static and flow-induced deformations around an isolated particle for $\text{\text{Bo}}\gg1$ and  $r \geq 1$ are given by (see Eqs. \eqref{eq:G2},\eqref{eq:G6}
\begin{align}
\text{Ca} h^{(1,0)}_{1p} \sim \frac{\text{Ca} 3\beta\cos\varphi}{2\text{Bo}r^4(3+\beta)}, \quad \text{ and } \delta h_{1p}^{(0,1)} \sim \frac{- \delta \tilde{\theta_s} e^{-\sqrt{\text{Bo}}(r-1)} }{\sqrt{\text{Bo} r }}.
\end{align}
This implies that for $(r,\varphi) \in \Sigma_{EFMN}^{(0,0)}$,  $h_{2p} \approx \text{Ca} h^{(1,0)}_{2p} + \delta h^{(0,1)}_{2p}$, where 
\begin{align}
    h^{(1,0)}_{2p}\approx \frac{3\beta\cos\varphi_1}{2\text{Bo}r_1^4(3+\beta)} + \frac{3\beta\cos\varphi_2}{2\text{Bo}r_2^4(3+\beta)},\quad \text{ and }h^{(0,1)}_{2p}\approx  \frac{-\tilde{\theta_s} e^{-\sqrt{\text{Bo}}(r-1)} }{\sqrt{\text{Bo} r }}  + \frac{-\tilde{\theta_s} e^{-\sqrt{\text{Bo}}(r_2-1)} }{\sqrt{\text{Bo} r_2 }}.
\end{align}
The surface integral in  Eq. \eqref{eqn:total_int_force:sum_capillary_viscous_forces} can now be evaluated, which gives 
\begin{align}
 -\frac{\text{Bo}}{\text{Ca}}  \iint_{\Sigma_{EFMN}^{(0,0)}} \text{Ca}  h_{2p}^{(1,0)}\mbox{ d}\Sigma  = &    -\frac{\text{Bo}}{\text{Ca}}  \iint_{\Sigma_{EFMN}^{(0,0)}} \text{Ca}\left( \frac{3\beta\cos\varphi_1}{2\text{Bo}r_1^4(3+\beta)} + \frac{3\beta\cos\varphi_2}{2\text{Bo}r_2^4(3+\beta)} \right)\mbox{ d}\Sigma  \\
 \approx & \frac{8\beta\cos\Phi}{L^2 (3+\beta)} -  \frac{3\pi\beta\cos\Phi}{2 L^4 (3+\beta)} 
\end{align}
and 
\begin{align}
 -\frac{\text{Bo}}{\text{Ca}}  \iint_{\Sigma_{EFMN}^{(0,0)}} \delta h^{(0,1)}_{2p}\mbox{ d}\Sigma = &  -\frac{\text{Bo}}{\text{Ca}}  \iint_{\Sigma_{EFMN}^{(0,0)}} \delta \left( \frac{-\tilde{\theta_s} e^{-\sqrt{\text{Bo}}(r_1-1)} }{\sqrt{\text{Bo} r_1 }}  + \frac{-\tilde{\theta_s} e^{-\sqrt{\text{Bo}}(r_2-1)} }{\sqrt{\text{Bo} r_2 }} \right)\mbox{ d}\Sigma \\
 \approx &  \frac{\delta 2\pi \tilde{\theta}_s}{\text{Ca}} . 
\end{align}
Note that the contribution from $\delta h_{2p}^{(0,1)}$ cancels out the capillary force contribution in the single-particle total force [see Eq. \eqref{eqn:total_int_force:F1p_solution}], i.e,
\begin{align}
    \frac{\delta}{\text{Ca}} 2\pi\tilde{\theta}_s \sqrt{\text{Bo}} C_0 K_1 (\sqrt{\text{Bo}}) \sim \frac{\delta 2\pi \tilde{\theta}_s}{\text{Ca}}
\end{align}
for $\text{Bo}\gg1.$
The vertical total interaction force is then given  by
\begin{align}
    \hat{\mathbf{e}}_z \cdot  \mathbf{F}_{tot}  =   \frac{192\beta \cos\Phi}{L^4 \text{Bo} (3+\beta) }  -  \frac{3\pi\beta\cos\Phi}{2 L^4 (3+\beta)},
\end{align}
which decays as $L^{-4}$ for $L \gg 1.$

\section{Conclusions}
In this work, we examined the translational motion of a spherical particle at a gas-liquid interface with surface slip, which extended the authors' analysis in Ref. \cite{zhou2022} that considered the problem of uniform flow past a spherical particle attached to an interface with a no-slip surface condition. The primary goal of this study is to understand the influence of interfacial deformations and the presence of surface slip on the hydrodynamic and interaction forces and torques exerted on a particle trapped at an interface. We adopted the two-parameter asymptotic approach used in Ref. \cite{zhou2022} where the interface shape is perturbed from the planar interface by assuming the Capillary number and the contact angle's deviation from 90 degrees are small. We analytically calculated the correction deformation, Ca$h^{(1,0)}$ , due to particle translation from O'Neill et al.'s zeroth order stress.

To obtain the drag and torque coefficients of a translating particle at a gas-liquid interface, we followed the asymptotic approach used in our earlier work, where the Lorentz reciprocal theorem was applied to calculate the correction drag and torque due to the interfacial deformation. To order Ca, drag and torque vanish because of the anti-symmetric shape of the flow-induced deformations, and the order $\delta$ drag and torque are computed analytically in terms of the $\mathcal{O}(1)$ flow field obtained by Ref. \cite{oneill1986} and the $\mathcal{O}(\delta)$ static deformation obtained in Ref. \cite{zhou2022}. Generally, the drag and torque coefficients exhibit a linear dependence on the correction contact angle and a nonlinear dependence on both the Bond number and the slip coefficient. 
For a particle translating along a deformable liquid interface, the analytical expression of the $\mathcal{O}(\delta)$ drag coefficient $k_D^{(0,1)}$ is given in terms of the $\mathcal{O}(\delta)$  translational solutions in Eq. \eqref{eqn:trans:orderDeltaDrag}. In the limit $\beta\rightarrow \infty$, Eq. \eqref{eqn:trans:orderDeltaDrag} recovers the drag coefficient obtained in Ref. \cite{zhou2022} without surface slip, and we find that $k_D^{(0,1)}$'s deviation from the no-slip solution decays as $\beta^{-1}$. In the limit $\beta\rightarrow 0$, the asymptotic expansion of  $k_D^{(0,1)}$  is given in Eq. \eqref{eqn:trans:perfect_slip_limit}, where the $\mathcal{O}(\beta^0)$ term recovers the perfect-slip solution. We also obtained the large-Bo approximations for $k_D^{(0,1)}$ [see Eq. \eqref{eqn:trans:large-Bo_drag_approx}], and for $\text{Bo}\gg 1,$  $k_D^{(0,1)}$ decays as $\text{Bo}^{-1/2}$. The $\mathcal{O}(\delta)$ torque coefficient of a translating particle,  $k_T^{(0,1)}$, is obtained by applying the Lorentz reciprocals theorem to the $\mathcal{O}(1)$ rotational flow field and the $\mathcal{O}(\delta)$ translational flow field [see Eqs. \eqref{eqn:trans:torque_T^t(0,1)_formula}  and \eqref{eqn:trans:torque_coeff_def}]. In the no-slip limit ($\beta\rightarrow\infty$),  $k_T^{(0,1)}$ is found to converge to a finite value monotonically for $\text{Bo} < 96.5.$ and nonmonotonically for $\text{Bo} \geq 96.5$. Moreover, we observed a similar decreasing Bo-dependence of $k_T^{(0,1)}$ to that of the drag coefficient $k_D^{(0,1)}$. In the limit of large Bond number, $k_T^{(0,1)}$ decays as $\text{Bo}^{-1/2}$.

A natural next step of our current work is to study the effect of the interfacial deformation on the hydrodynamic force and torque exerted on a rotating particle at a liquid interface. However, due to the singular behavior of the $\mathcal{O}(1)$ rotational stress $\bm{\sigma}^{r(0,0)}$ near the TCL, the analytical calculations of the drag and torque coefficients pose a nontrivial challenge. The analysis required to account for the complex boundary layer dynamics near the TCL goes beyond the scope of our current perturbation method.

Additionally, we studied the capillary interactions between two translating and rotating particles at a large separation. Paralleling D\"orr and Hardt's analysis in Ref. \cite{dorr2015}, we constructed the interfacial deformation around two particles from the single-particle deformation via LSA and computed the capillary interaction force and torque exerted on the translating and rotating particles. Note that the rotational-flow-induced deformation Ca $h^{r(1,0)}$ is analytically calculated from the zeroth order rotational stress in Ref. \cite{oneill1986} and given 
in Appendix \ref{App:rot-flow-induced-deformation} . 
Following Ref. \cite{dorr2015}'s multipole analysis, the flow-induced deformations \eqref{eqn:flow_induced_deformations_solutions} are identified as dipolar in terms of the azimuthal angle $\varphi$ and the interaction forces \eqref{eqn:trans:cap_force_formula_2} and \eqref{eqn:rot:cap_force_formula_2} due to the flow-induced deformations as dipolar forces. In the limit of large separation ($L\rightarrow\infty$), the lateral interaction force in \eqref{eqn:trans:cap_force_formula_2} and \eqref{eqn:rot:cap_force_formula_2} decays as $L^{-9}$, while the vertical interaction forces decay as $L^{-4}.$ The decay rates of the interaction forces are higher than those of the interaction forces obtained in Ref. \cite{dorr2015}. The difference is due to the effect of the gravity forces which flattens the gas-liquid interface and speeds up the decay.  Finally we have estimated the total vertical interaction force on one of the two translating particles in the large separation and large Bond number limit.  We find the force decays as $1/L^4$ and can be positive or negative depending on the location relative to the applied flow.  Future work will need to address these estimates beyond the asymptotoc limiting assumptions used here.

\begin{acknowledgments}
This work was partially supported by NSF grant DMS 2108502. 
\end{acknowledgments}

\newpage
\section*{Appendices}
\appendix

\section{Comparison of linear and nonlinear static deformations}\label{app:comparison_2D/3D_interfaces}

\begin{figure}
    \centering
    \includegraphics[width=5.5cm]{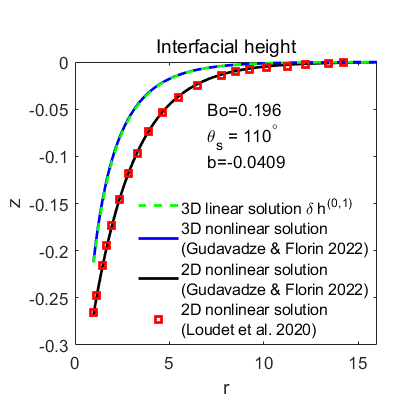}
    \caption{Comparison of the 3D linear static deformation \eqref{eqn:static_deformation} with the numerical solutions of 2D and 3D nonlinear interfacial shapes with Bond number $\text{Bo} = 0.196,$ contact angle $\theta_s = 110^\circ$, and immersion depth $b = -0.0409.$  }
    \label{fig:app:comparison_2D/3D_interfaces_loudet}
\end{figure}

\begin{figure}
    \centering
    \includegraphics[width=11cm]{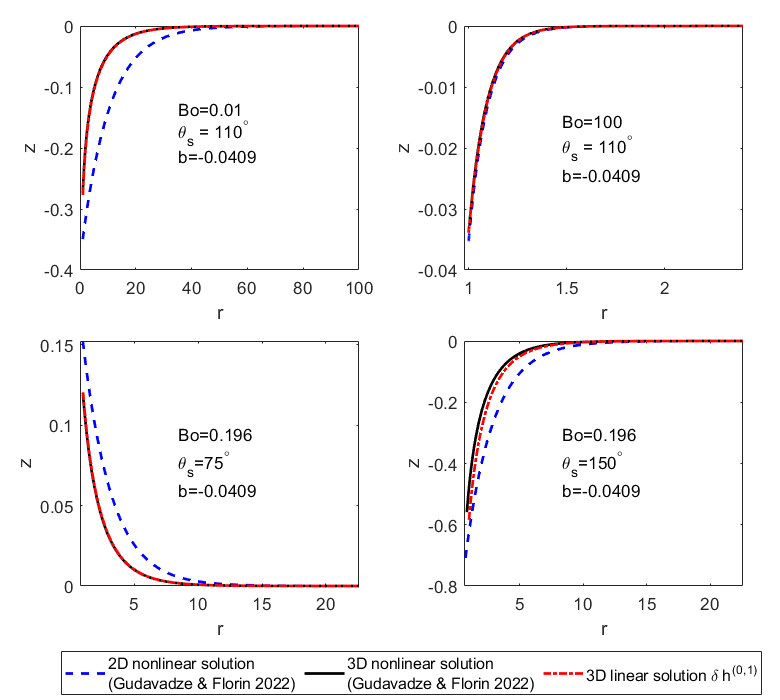}
 \caption{Comparison of the 3D linear static deformations \eqref{eqn:static_deformation} with the numerical solutions of 2D and 3D nonlinear interfacial shapes with Bond number $\text{Bo} = 0.01, 0.196$ and 100, contact angle $\theta_s = 75^\circ, 110^\circ$ and $150^\circ$, and immersion depth $b = -0.0409.$}
    \label{fig:app:comparison_2D/3D_interfaces_Bo_angle}
\end{figure}

Following Gudavadze and Florin  \cite{Gudavadze2022} the 3D nonlinear static interface shape around a floating particle can be  numerically determined by solving the Young-Laplace equation using MATLAB's BVP5C solver \cite{Gudavadze2022}. We have implemented this numerical approach in both 3D and 2D.   In Fig. \ref{fig:app:comparison_2D/3D_interfaces_loudet}, we compare the 3D linear static interface shape $\delta h^{(0,1)}$ in Eq. \eqref{eqn:static_deformation} with the 2D and 3D nonlinear interface shapes with $\text{Bo} = 0.196,$ contact angle $\theta_s = 110^{\circ}$, and immersion depth $b = -0.0409.$ We discover that the 3D linear approximation $\delta h^{(0,1)}$  agrees very well with the 3D nonlinear results, while quantitative differences are observed between the 2D and 3D solutions. In Figs. \ref{fig:app:comparison_2D/3D_interfaces_Bo_angle}, we plot $\delta h^{(0,1)}$ along with the 2D and 3D nonlinear interface shapes with varying values of Bond number and contact angle $\theta_s.$ 
For $0.01 \geq \text{Bo} \geq 100$ and $75^\circ  \geq \theta_s \geq 150$, the 3D linear approximation $\delta h^{(0,1)}$ shows good agreement with the 3D nonlinear results, and differs when compared to the 2D solution.
 This implies that the model's dimensionality plays a significant role in the equilibrium static interface shape. For $\text{Bo} = 100,$ the gravity force exerted on the liquid phase dominates the dimensionality effect and we see that the 3D and 2D interface shapes start to overlap. 

Note that in our earlier work \cite{zhou2022}, the analytical calculations of drag coefficients of a translating sphere with no-slip boundary conditions agreed well qualitatively but not quantitatively with the 2D full flow solutions by Loudet \textit{et al.} \cite{Loudet2020} for $\text{Bo} = 0.196$ and $\theta_s = 75^\circ$ and $110^\circ$. Since the equilibrium static interface shape has a significant impact on the deformation-induced drag force, the comparisons in Figs. \ref{fig:app:comparison_2D/3D_interfaces_loudet} and \ref{fig:app:comparison_2D/3D_interfaces_Bo_angle} suggest that the quantitative differences observed in  Ref. \cite{zhou2022} can be attributed to the models' different dimensionalities and static interface shapes. Also, the excellent agreement between $\delta h^{(0,1)}$ and the nonlinear static deformation for contact angle $\theta_s$ ranging from $75^\circ$ to $150^\circ$ implies that the asymptotic models in Ref. \cite{zhou2022} and our current work are robust to large contact angle deviations. However, further numerical verifications are still required to fully understand the effect of contact angles.

\section{The $\mathcal{O}(1)$ velocity field and pressure} \label{App:order1_solution}
For notational convenience, in this section we denote the translational velocity field $\mathbf{u}$ and pressure $p$ as $\mathbf{u}^t$ and $p^t$, respectively. The $\mathcal{O}(1)$ velocity field $\mathbf{u}^{t,r(0,0)} = u^{t,r(0,0)}_\rho \hat{\mathbf{e}}_\rho + u^{t,r(0,0)}_\theta \hat{\mathbf{e}}_\theta + u^{t,r(0,0)}_\phi \hat{\mathbf{e}}_\phi$ and pressure $p^{t,r(0,0)}$ obtained by \cite{oneill1986} are given by 
 \begin{align}
 &
    u^{t,r(0,0)}_\rho =    \cos\phi \sum_{n=0}^\infty (2n+1)(2n+2)\left(\frac{A_n}{\rho^{2n+3}} + \frac{B_n}{\rho^{2n+1}}\right) \sin\theta P'_{2n+1},\label{eqn:app:order1_sol1}
  \\ & u^{t,r(0,0)}_\theta  =  \cos\phi \sum_{n=0}^\infty \left [\left( \frac{(2n+1)A_n}{\rho^{2n+3}} + \frac{(2n-1)B_n}{\rho^{2n+1}} \right) \left( \frac{(2n+2)^2}{4n+3} P'_{2n} \right. \right. \nonumber  \\  & - \left. \left.\frac{(2n+1)^2}{4n+3} P'_{2n+2} \right)  + \frac{C_n}{\rho^{2n+1}} P'_{2n} \right], \label{eqn:app:order1_sol2}
    \\ 
    & u^{t,r(0,0)}_\phi = \sin\phi \sum_{n=0}^\infty \left[\left( \frac{(2n+1)A_n}{\rho^{2n+3}} + \frac{(2n-1)B_n}{\rho^{2n+1}} \right)  P'_{2n+1} \right.  \nonumber \\ & +\left. \frac{C_n}{\rho^{2n+1}} \left( \frac{(2n+1)^2}{4n+1} P'_{2n-1} - \frac{(2n)^2}{4n+1} P'_{2n+1}\right) \right],\label{eqn:app:order1_sol3}\\
    & p^{t,r(0,0)}  =  \frac{\cos\phi}{\sin\theta} \frac{\partial}{\partial \rho} \left[ \frac{\partial^2 \psi }{\partial\rho^2} + \frac{\sin\theta}{\rho} \frac{\partial}{\partial\theta} \left( \frac{1}{\sin\theta} \frac{\partial\psi}{\partial \theta} \right) \right], \label{eqn:app:order1_sol4} 
\end{align}
where $P_n$ is the Legendre polynomial of order $n$ with argument $\cos\theta,$  $A_n, B_n,$ and $C_n$ are coefficients dependent on the slip coefficient $\beta,$ and $\psi$ is defined by 
\begin{align}
    \psi = \sum_{n=0}^\infty \left( \frac{A_n}{\rho^{2n+1}} + \frac{B_n}{\rho^{2n-1}} \right)P'_{2n+1} \sin^2\theta. 
 \end{align}
 In the $\mathcal{O}$(1) problem of a translating particle bisected by a flat gas-liquid interface, the coefficients $A_n, B_n$ and $C_n$ are determined to be
\begin{align}
    A_0 = -\frac{1}{4} \frac{\beta}{\beta+3}, \quad B_0 = \frac{3}{4} \frac{\beta+2}{\beta+3}, \quad C_0 = 0,
\end{align} 
and $A_n = B_n = C_n = 0$ for $n \geq 1. $ In the $\mathcal{O}(1)$ problem of a rotating particle, $A_n, B_n$ and $C_n$ are expressed in terms of coefficients $T_n$ and $S_n$, 
\begin{align}
    A_n = -B_n = \frac{\beta S_n}{2(3+4n+\beta)}, \quad C_n = \frac{\beta T_n}{2+2n+\beta},
\end{align} and  $S_n$ and $T_n$ are be determined from the following recurrence formula, i.e.,
\begin{align}
&  T_0 = 0, \quad S_0 = -\frac{3}{4}, \quad S_1 = \frac{7}{16}\left(\frac{S_0}{3} - T_1 + \alpha_1 \right),\\
&     T_n =   \frac{(4n+1)}{(2n+1)^2} \sum_{n=0}^{n-1} (\beta_i - S_i - T_i), \quad n\geq1,\\
 &    S_n  = \frac{(4n+3)}{(2n+2)^2} \sum_{n=1}^{n-1} \left( (\alpha_i - S_i - T_i) + \alpha_n- T_n -\frac{S_0}{3} \right), \quad n\geq2 ,   \end{align}
with
\begin{align*}
    & \alpha_n = -\frac{(4 n +1)t_n}{2}, \quad \beta_n = \frac{(4n+3)t_n}{4(n+2)}, \quad t_n = \frac{P_{2n}(0)}{(n+1)(2n-1)}.
\end{align*}

\section{The Lorentz reciprocal theorem}\label{app:reciprocal_theorem}
\subsection{Derivation of the hydrodynamic force}\label{app:reciprocal_theorem:drag}
Substituting the asymptotic expansions of $\mathbf{u}^{t}$ and $\bm{\sigma}^t$ into Eq. \eqref{eqn:trans:reciprocal_theorem_relation} and collecting coefficients of Ca, we obtain 
{   
\begin{align}
\begin{split}
& \iint_{\Sigma_{P}^{(0,0)}} (\bm{\sigma}^{t(0,0)}\cdot (-\hat{\mathbf{e}}_\rho) \cdot ( \mbox{Ca} \mathbf{u}^{t(1,0)} + \delta \mathbf{u}^{t(0,1)} ) \mbox{ d}\Sigma  \\ & + \iint_{\Sigma_{I^{(0,0)}}} \bm{\sigma}^{t(0,0)}  \cdot (- \hat{\mathbf{e}}_z)   \cdot \left(  \mbox{Ca}    \mathbf{u}^{t(1,0)}   + \delta \mathbf{u}^{t(0,1)}\right)  \mbox{ d}\Sigma  \\
= & \iint_{\Sigma_{P}^{(0,0)}} (\mbox{Ca}  \bm{\sigma}^{t(1,0)} + \delta \bm{\sigma}^{t(0,1)} ) \cdot (-\hat{\mathbf{e}}_\rho)  \cdot  \mathbf{u}^{t(0,0)} \mbox{ d}\Sigma  \\ &  + \iint_{\Sigma_{I^{(0,0)}}} \left( \mbox{Ca}   \bm{\sigma}^{t(1,0)} + \delta \bm{\sigma}^{t(0,1)} \right)  \cdot (- \hat{\mathbf{e}}_z) \cdot \mathbf{u}^{t(0,0)} \mbox{ d}\Sigma,
\end{split} 
\label{eqn:app:reciprocal_eqn1}\end{align}
}
Rearranging terms in  Eq. \eqref{eqn:app:reciprocal_eqn1} yields   
{ 
\begin{align} &
\iint_{\Sigma_{P}^{(0,0)}} ( \mbox{Ca}  \bm{\sigma}^{t(1,0)} + \delta \bm{\sigma}^{t(0,1)} ) \cdot \hat{\mathbf{e}}_\rho  \cdot \mathbf{V}^t  \mbox{ d}\Sigma \\ &
     =  -\iint_{\Sigma_{P}^{(0,0)}} ( \mbox{Ca}  \bm{\sigma}^{t(1,0)} + \delta \bm{\sigma}^{t(0,1)} ) \cdot \hat{\mathbf{e}}_\rho\cdot (\mathbf{u}^{t(0,0)}-\mathbf{V}^t ) \mbox{ d}\Sigma  \label{eqn:app:reciprocal_eqn2_1}\\
      & - \iint_{\Sigma_{I^{(0,0)}}} \left( \mbox{Ca}   \bm{\sigma}^{t(1,0)} + \delta \bm{\sigma}^{t(0,1)} \right)  \cdot \hat{\mathbf{e}}_z \cdot \mathbf{u}^{t(0,0)} \mbox{ d}\Sigma  \label{eqn:app:reciprocal_eqn2_2}\\
     & + \iint_{\Sigma_{P}^{(0,0)}} (\bm{\sigma}^{t(0,0)}\cdot \hat{\mathbf{e}}_\rho\cdot ( \mbox{Ca} \mathbf{u}^{t(1,0)} + \delta \mathbf{u}^{t(0,1)} ) \mbox{ d}\Sigma  \label{eqn:app:reciprocal_eqn2_3} \\
& + \iint_{\Sigma_{I^{(0,0)}}} \bm{\sigma}^{t(0,0)}  \cdot \hat{\mathbf{e}}_z  \cdot \left(  \mbox{Ca}    \mathbf{u}^{t(1,0)}   + \delta \mathbf{u}^{t(0,1)}\right)  \mbox{ d}\Sigma. \label{eqn:app:reciprocal_eqn2_4}
\end{align}
}

Substituting Eqs. \eqref{eqn:trans:reciprocal_slip_condition_O(1)} and \eqref{eqn:trans:reciprocal_slip_condition} for $\mathbf{u}^{t(0,0)}$, $\mathbf{u}^{t(1,0)}$, and $\mathbf{u}^{t(0,1)}$, we find that the integrals \eqref{eqn:app:reciprocal_eqn2_1} and \eqref{eqn:app:reciprocal_eqn2_3} conveniently cancel out, i.e., 
{ 
\begin{align}
  & -\iint_{\Sigma_{P}^{(0,0)}}
  \mbox{Ca}  \bm{\sigma}^{t(1,0)} \cdot \hat{\mathbf{e}}_\rho\cdot (\mathbf{u}^{t(0,0)}-\mathbf{V}^t ) \mbox{ d}\Sigma  + \iint_{\Sigma_{P}^{(0,0)}} \bm{\sigma}^{t(0,0)}\cdot \hat{\mathbf{e}}_\rho \cdot  \mbox{Ca} \mathbf{u}^{t(1,0)}  \mbox{ d}\Sigma \nonumber  \\
  = &  -\iint_{\Sigma_{P}^{(0,0)}} \mbox{Ca} \left(\sigma^{t(1,0)}_{\rho \rho } \hat{\mathbf{e}}_\rho + \sigma^{t(1,0)}_{\rho\theta}\hat{\mathbf{e}}_\theta + \sigma^{t(1,0)}_{\rho\phi} \hat{\mathbf{e}}_\phi \right)\cdot \frac{1}{\beta}\left(\sigma^{t(0,0)}_{\rho \theta} \hat{\mathbf{e}}_{\theta} + \sigma^{t(0,0)}_{\rho \phi} \hat{\mathbf{e}}_\phi \right) \mbox{ d}\Sigma \nonumber \\
  & +  \iint_{\Sigma_{P}^{(0,0)}}  \left(\sigma^{t(0,0)}_{\rho \rho} \hat{\mathbf{e}}_\rho + \sigma^{t(0,0)}_{\rho\theta}\hat{\mathbf{e}}_\theta + \sigma^{t(0,0)}_{\rho\phi} \hat{\mathbf{e}}_\phi \right)\cdot \frac{\mbox{Ca}}{\beta}\left(\sigma^{t(1,0)}_{\rho \theta} \hat{\mathbf{e}}_\theta + \sigma^{t(1,0)}_{\rho \phi} \hat{\mathbf{e}}_\phi\right) \mbox{ d}\Sigma  \nonumber \\
  = & -\iint_{\Sigma_{P}^{(0,0)}} \frac{\mbox{Ca}}{\beta} \left( \sigma_{\rho \theta}^{t(1,0)} \sigma_{\rho \theta }^{t(0,0)} + \sigma_{\rho \phi}^{t(1,0)} \sigma_{\rho \phi }^{t(0,0)}  \right) \mbox{ d}\Sigma \nonumber \\ & + \iint_{\Sigma_{P}^{(0,0)}} \frac{\mbox{Ca}}{\beta} \left( \sigma_{\rho \theta}^{t(1,0)} \sigma_{\rho \theta }^{t(0,0)} + \sigma_{\rho \phi}^{t(1,0)} \sigma_{\rho \phi }^{t(0,0)}  \right) \mbox{ d}\Sigma \nonumber \\
   = & 0. \nonumber
  \end{align}}

Note that similar relation holds for the $\mathcal{O}(\delta)$ terms. Then, Eqs. \eqref{eqn:app:reciprocal_eqn2_1} - \eqref{eqn:app:reciprocal_eqn2_4} and their $\mathcal{O}(\delta)$ counterparts are reduced to
\begin{align}
 &    \iint_{\Sigma_{P}^{(0,0)}} \bm{\sigma}^{t(1,0)} \cdot \hat{\mathbf{e}}_\rho  \cdot \mathbf{V}^t  \mbox{ d}\Sigma = \iint_{\Sigma_I^{(0,0)}} \bm{\sigma}^{t(0,0)}\cdot \hat{\mathbf{e}}_z\cdot \mathbf{u}^{t(1,0)} \mbox{ d}\Sigma -  \iint_{\Sigma_I^{(0,0)}} \bm{\sigma}^{t(1,0)}\cdot \hat{\mathbf{e}}_z\cdot \mathbf{u}^{t(0,0)} \mbox{ d}\Sigma,\label{eqn:app:reciprocal_eqn3_1}\\
 &   \iint_{\Sigma_{P}^{(0,0)}} \bm{\sigma}^{t(0,1)} \cdot \hat{\mathbf{e}}_\rho \cdot \mathbf{V}^t  \mbox{ d}\Sigma = \iint_{\Sigma_I^{(0,0)}} \bm{\sigma}^{t(0,0)}\cdot \hat{\mathbf{e}}_z\cdot \mathbf{u}^{t(0,1)} \mbox{ d}\Sigma -  \iint_{\Sigma_I^{(0,0)}} \bm{\sigma}^{t(0,1)}\cdot \hat{\mathbf{e}}_z\cdot \mathbf{u}^{t(0,0)} \mbox{ d}\Sigma. \label{eqn:app:reciprocal_eqn3_2}
\end{align}

\subsection{Derivation of the hydrodynamic torque}\label{app:reciprocal_theorem:torque}
Substituting the asymptotic expansions of $\mathbf{u}^t$ and $\bm{\sigma}^t$ into Eq. \eqref{eqn:rot:reciprocal_theorem_relation}, collecting coefficients of Ca and rearranging terms, we arrive at  
{ 
\begin{align}
 & \iint_{\Sigma_{P}^{(0,0)}} ( \mbox{Ca}  \bm{\sigma}^{t(1,0)} + \delta \bm{\sigma}^{t(0,1)} ) \cdot \hat{\mathbf{e}}_\rho  \cdot \mathbf{V}^r  \mbox{ d}\Sigma 
     \\ & =  -\iint_{\Sigma_{P}^{(0,0)}} ( \mbox{Ca}  \bm{\sigma}^{t(1,0)} + \delta \bm{\sigma}^{t(0,1)} ) \cdot \hat{\mathbf{e}}_\rho\cdot (\mathbf{u}^{r(0,0)}-\mathbf{V}^r ) \mbox{ d}\Sigma  \label{eqn:app:torque:reciprocal_eqn2_1}\\
      & - \iint_{\Sigma_{I^{(0,0)}}} \left( \mbox{Ca}   \bm{\sigma}^{t(1,0)} + \delta \bm{\sigma}^{t(0,1)} \right)  \cdot \hat{\mathbf{e}}_z \cdot \mathbf{u}^{r(0,0)} \mbox{ d}\Sigma  \label{eqn:app:torque:reciprocal_eqn2_2}\\
     & + \iint_{\Sigma_{P}^{(0,0)}} (\bm{\sigma}^{r(0,0)}\cdot \hat{\mathbf{e}}_\rho\cdot ( \mbox{Ca} \mathbf{u}^{t(1,0)} + \delta \mathbf{u}^{t(0,1)} ) \mbox{ d}\Sigma  \label{eqn:app:torque:reciprocal_eqn2_3} \\
& + \iint_{\Sigma_{I^{(0,0)}}} \bm{\sigma}^{r(0,0)}  \cdot \hat{\mathbf{e}}_z  \cdot \left(  \mbox{Ca}    \mathbf{u}^{t(1,0)}   + \delta \mathbf{u}^{t(0,1)}\right)  \mbox{ d}\Sigma. \label{eqn:app:torque:reciprocal_eqn2_4}
\end{align}
}

Substituting Eq. \eqref{eqn:trans:reciprocal_slip_condition} for $\mathbf{u}^{t(1,0)}$ and $\mathbf{u}^{t(0,1)}$ and Eqs. \eqref{eqn:rot:navier_slip_cond_theta} and \eqref{eqn:rot:navier_slip_cond_phi} for $\mathbf{u}^{r(0,0)}$, we show that the integrals \eqref{eqn:app:torque:reciprocal_eqn2_1} and \eqref{eqn:app:torque:reciprocal_eqn2_3} also conveniently cancel out, i.e., 
{ 
\begin{align}
  & -\iint_{\Sigma_{P^{(0,0)}}}
  \mbox{Ca}  \bm{\sigma}^{t(1,0)} \cdot \hat{\mathbf{e}}_\rho\cdot (\mathbf{u}^{r(0,0)}-\mathbf{V}^r ) \mbox{ d}\Sigma  + \iint_{\Sigma_{P}^{(0,0)}} \bm{\sigma}^{r(0,0)}\cdot \hat{\mathbf{e}}_\rho \cdot  \mbox{Ca} \mathbf{u}^{t(1,0)}  \mbox{ d}\Sigma  \nonumber \\
  = &  -\iint_{\Sigma_{P}^{(0,0)}} \mbox{Ca} \left(\sigma^{t(1,0)}_{\rho \rho } \hat{\mathbf{e}}_\rho + \sigma^{t(1,0)}_{\rho\theta}\hat{\mathbf{e}}_\theta + \sigma^{t(1,0)}_{\rho\phi} \hat{\mathbf{e}}_\phi \right)\cdot \frac{1}{\beta}\left(\sigma^{r(0,0)}_{\rho \theta} \hat{\mathbf{e}}_{\theta} + \sigma^{r(0,0)}_{\rho \phi} \hat{\mathbf{e}}_\phi \right) \mbox{ d}\Sigma \nonumber \\
  & +  \iint_{\Sigma_{P}^{(0,0)}}  \left(\sigma^{r(0,0)}_{\rho \rho} \hat{\mathbf{e}}_\rho + \sigma^{r(0,0)}_{\rho\theta}\hat{\mathbf{e}}_\theta + \sigma^{r(0,0)}_{\rho\phi} \hat{\mathbf{e}}_\phi \right)\cdot \frac{\mbox{Ca}}{\beta}\left(\sigma^{t(1,0)}_{\rho \theta} \hat{\mathbf{e}}_\theta + \sigma^{t(1,0)}_{\rho \phi} \hat{\mathbf{e}}_\phi\right) \mbox{ d}\Sigma \nonumber \\
  = & -\iint_{\Sigma_{P}^{(0,0)}} \frac{\mbox{Ca}}{\beta} \left( \sigma_{\rho \theta}^{t(1,0)} \sigma_{\rho \theta }^{r(0,0)} + \sigma_{\rho \phi}^{t(1,0)} \sigma_{\rho \phi }^{r(0,0)}  \right) \mbox{ d}\Sigma \nonumber \\ & + \iint_{\Sigma_{P}} \frac{\mbox{Ca}}{\beta} \left( \sigma_{\rho \theta}^{t(1,0)} \sigma_{\rho \theta }^{r(0,0)} + \sigma_{\rho \phi}^{t(1,0)} \sigma_{\rho \phi }^{r(0,0)}  \right) \mbox{ d}\Sigma \nonumber \\
   = & 0. \nonumber
  \end{align}}
Similar relation holds for the $\mathcal{O}(\delta)$ terms, and Eqs. \eqref{eqn:app:torque:reciprocal_eqn2_1} - \eqref{eqn:app:torque:reciprocal_eqn2_4} and their $\mathcal{O}(\delta)$ counterparts are simplified to be
\begin{align}
 &  T^{(1,0)} =   \iint_{\Sigma_{P}^{(0,0)}} \bm{\sigma}^{t(1,0)} \cdot \hat{\mathbf{e}}_\rho  \cdot \mathbf{V}^r  \mbox{ d}\Sigma \nonumber  \\ & = \iint_{\Sigma_I^{(0,0)}} \bm{\sigma}^{r(0,0)}\cdot \hat{\mathbf{e}}_z\cdot \mathbf{u}^{t(1,0)} \mbox{ d}\Sigma -  \iint_{\Sigma_I^{(0,0)}} \bm{\sigma}^{t(1,0)}\cdot \hat{\mathbf{e}}_z\cdot \mathbf{u}^{r(0,0)} \mbox{ d}\Sigma,\label{eqn:app:torque:reciprocal_eqn3_1}\\
 & T^{(0,1)} =   \iint_{\Sigma_{P}^{(0,0)}} \bm{\sigma}^{t(0,1)} \cdot \hat{\mathbf{e}}_\rho \cdot \mathbf{V}^t  \mbox{ d}\Sigma \nonumber \\ & = \iint_{\Sigma_I^{(0,0)}} \bm{\sigma}^{r(0,0)}\cdot \hat{\mathbf{e}}_z\cdot \mathbf{u}^{t(0,1)} \mbox{ d}\Sigma  -  \iint_{\Sigma_I^{(0,0)}} \bm{\sigma}^{t(0,1)}\cdot \hat{\mathbf{e}}_z\cdot \mathbf{u}^{r(0,0)} \mbox{ d}\Sigma. \label{eqn:app:torque:reciprocal_eqn3_2}
\end{align}

\section{Comparison with literature results}
\begin{figure}
    \centering
    \includegraphics[scale=0.6]{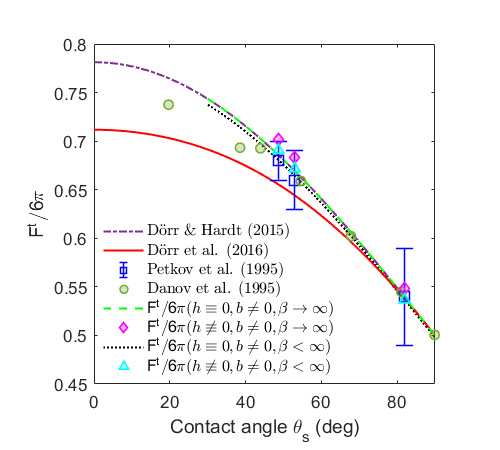}
    \caption{Comparison of the normalized drag $F^t/6\pi$ computed from Eq. \eqref{eqn:app:translational_drag_formula} with experimental and theoretical results from Refs. \cite{Danov1995, dorr2015, dorr2016, Petkov1995}.  }
    \label{fig:app:literature_comparison}
\end{figure}

Here, we would like to compare our drag force calculation \eqref{eqn:trans:F_t_formula2} of a particle translating along a gas-liquid interface with the experimental and theoretical results from the literature. In the experimental study \cite{Petkov1995}, Petkov \textit{et al.} determined the drag coefficients of isolated hydrophobized submillimeter glass beads at an air-water interface, where the particle translation was driven by the the controlled capillary interaction between the glass bead and a vertical wall. The drag coefficient was calculated by comparing the numerically simulated particle motion and the measured particle trajectory, and the equilibrium contact angle was determined from the measured immersion depth or the radius of the TCL using a iterative procedure described by Eqs. (3.1)--(3.9) in Ref. \cite{Velev1994}. In the theoretical predictions by Danov \textit{et al.} \cite{Danov1995}, D\"orr and Hardt \cite{dorr2015}, and D\"orr \textit{et al.} \cite{dorr2016}, the drag coefficients were calculated under the assumptions that the particle is allowed to shift vertically with an immersion depth $b$, and that the gas-liquid interface does not deform around the particle, in which case, the contact angle $\theta_s$ and the immersion depth $b$ (scaled by $a$) satisfy the relation $\cos\theta_s = b$.  In our earlier work \cite{zhou2022}, we computed the drag coefficient of a translating particle with no-slip surface conditions and immersion depth $b = \delta \tilde{b}$. In our current model with finite slip ($\beta <\infty$), the $\mathcal{O}(\delta)$ drag contribution from the particle's vertical displacement $\delta \tilde{b}$ can be easily obtained by paralleling the analysis in Ref. \cite{zhou2022}, which gives us the asymptotic approximation of the translational drag  $F^{} = F^{(0,0)} + \delta F^{(0,1)}$, where the $\mathcal{O}(1)$ drag is given by 
{\small 
\begin{align}
\begin{split}
 F^{(0,1)}  = &  \int_1^\infty \frac{9\pi \left((\beta +2) r^2-\beta
   \right)  }{8 (\beta
   +3)^2 r^7}  \left[-6 (\beta +2) r^2
  h^{(0,1)}+6 (\beta +2) r^3
  \frac{\partial  h^{(0,1)}}{\partial r}+4 \beta   h^{(0,1)} \right. \\ & -3
   \left. \beta  r \frac{\partial  h^{(0,1)}}{\partial r}\right] \mbox{ d}r  -
    3\pi (\Tilde{\theta}_s -\tilde{b})C_0 K_0(\sqrt{\text{Bo}}) - \frac{  9 \pi \tilde{b}(18+7\beta + 3\beta^2)}{16(2+\beta)^2},
   \end{split}  \label{eqn:app:translational_drag_formula}
\end{align}}
and the $\mathcal{O}(\delta)$ static deformation $h^{(0,1)}$ is given in Eq. \eqref{eqn:static_deformation}. The numerical values of $b$ is determined from the particle radius $a$ and the equilibrium contact angle $\theta_s$ by using the iterative procedure described by Eqs. (3.1)--(3.9) in Ref. \cite{Velev1994}. For particles with different radii $a$ and contact angles $\theta_s$, the numerical values of immersion depth $b$ and Bond number Bo used in the force calculations are given in Table \ref{tab:app:parameter_values}. 

\begin{table}[b]
\caption{\label{tab:app:parameter_values}%
Parameter values used in the drag force calculations in Fig. \ref{fig:app:literature_comparison} for comparison with the experimental results obtained by Petkov et al. \cite{Petkov1995}.
}
\begin{ruledtabular}
{\small
\begin{tabular}{ccccc}
\newcommand{\wrap}[1]{\parbox{.33\linewidth}{\vspace{1.5mm}#1\vspace{1mm}}}
{\shortstack{\textrm{Contact angle}\\ $\theta_s (^\circ)$}} &
{\shortstack{\textrm{Particle radius}\\ $a$ (mm)}}   &
\shortstack{\textrm{Immersion depth }$b$\\ (nondimensionalized)} &
{\shortstack{\textrm{Slip coefficient }$\beta$\\ (nondimensionalized)}} & {\shortstack{\textrm{Bond number}\\ Bo}} \\
\colrule
48 & 0.229 & -0.658 & 229 & 0.0072\\
53 & 0.231 & -0.600 & 231& 0.0073\\
82 & 0.222 & -0.142 & 222& 0.0068\\
\end{tabular}
}
\end{ruledtabular}
\end{table}

To recover the case of an undeformable flat interface, we can simply set $\tilde{\theta}_s = \tilde{b}$.  The no-slip and finite-slip drag coefficients with a flat interface and a nonzero immersion depth are shown as a dashed green curve and a dotted black curve, respectively, where the no-slip drag coefficient reproduces the D\"orr and Hardt's result from Ref. \cite{dorr2015} (purple dash-dotted curve).
In the case of particle translation along a deformable interface, we recover the immersion depth $b$ from the equilibrium contact angle $\theta_s$ calculated in Ref. \cite{Petkov1995} by applying the iterative procedure described in Ref. \cite{Velev1994}. For contact angle $\theta_s = 48^\circ, 53^\circ,$ and $82^\circ$, the no-slip and finite-slip drag coefficients accounting for interfacial deformation ($h \not\equiv 0$) and particle undulation ($b \neq 0$) are shown as magenta diamonds  and cyan triangles, respectively.  To determine the slip coefficient $\beta$, we use the slip length $s = 1 \mu m$ measured in Ref. \cite{Tretheway2002} for water flowing over a hydrophobic glass surface and set $\beta = s/a$, where $a$ is the radius of the glass bead. The values of slip coefficient $\beta$ used in the force calculations is given in Tab. \ref{tab:app:parameter_values}. 
In Fig \ref{fig:app:literature_comparison}, we reproduce the comparison of experimental and theoretical results in Fig. 3 from Ref. \cite{dorr2016} while including the drag predictions from our analytical models. 
As shown in Fig.  \ref{fig:app:literature_comparison}, 
our asymptotic approximations of the drag coefficients are consistent with the results taken from the literature. In particular, we observe that for $\theta_s = 48^\circ, 53^\circ,$ and $89^\circ$, our drag calculations with finite slip lengths ($s = 1 \mu m$) have a better agreement with the average values of Ref. \cite{Petkov1995}'s experimental measurements compared to their no-slip counterparts. However, further verification is still needed due to the complex nature of the physical systems of surface-trapped particles. 
\\

 \section{Behaviors of the static deformation for small and large Bond number} \label{App:asymptitics_bond_number}
Note that for $r \gg 1$ and $\text{Bo} \sim 1$, the static deformation  $\delta h^{(0,1)}$ behaves asymptotically as 
\begin{align}
   \delta  h^{(0,1)} \sim \sqrt{\frac{\pi}{2}}\frac{-\tilde{\theta}_s  e^{-\sqrt{\text{Bo}} r} }{(\sqrt{\text{Bo}} r)^{1/2}  [K_1(\sqrt{\text{Bo}}) + \sqrt{\text{Bo}} K_1(\sqrt{\text{Bo}})]}. 
\end{align}
 This implies that $\delta h^{(0,1)}$ away from the particle decays to zero as  $ \sqrt{\text{Bo}} r \rightarrow 0$. Therefore, as Bond number Bo decreases, $\delta h^{(0,1)}$ decays at a slower rate. For $\text{Bo} = 0, $ the static deformation $\delta h^{(0,1)} \equiv -\delta \Tilde{\theta}_s$, which fails to satisfy the far-field condition \eqref{eqn:far_field_condition_static_def} and becomes invalid. On the other hand, for Bo $\gg 0$, the gravity effect dominates the interfacial dynamics and $\delta h^{(0,1)}$ decays to zero at a much faster rate, i.e., 
  \begin{align}
      h^{(0,1)} \sim \frac{-\tilde{\theta}_s e^{-\sqrt{\text{Bo}}(1-r)}}{\sqrt{\text{Bo} r}}. 
  \end{align}

\section{Rotational-flow-induced deformation}\label{App:rot-flow-induced-deformation}
   The interfacial deformation induced by the particle's rotational motion can be obtained by paralleling the calculation in Sec. \ref{sec:interfacial_deformations}. 
Here, the leading-order normal-normal stress $\sigma_{zz}^{r(0,0)}$ at the interface $\Sigma_I^{(0,0)}$ is computed from the rotational flow field $\mathbf{u}^{r(0,0)}$ obtained by O'Neill \textit{et al.} \cite{oneill1986}, i.e., 
\begin{align}
     \sigma_{zz}^{r(0,0)}  =  F^r(r)\cos\varphi,
\end{align}
with 
    \begin{align}
    F^r(r) =  \sum_{n=0}^\infty -\frac{4\sqrt{\pi}  (A_n (2n+1)^2 (2n+3) + 2n(C_n +2 B_n (2n^2 + n-1) )  r^2 ) }{\Gamma(-1/2-n)\Gamma(n+1) r^{2(n+2)}}. 
\end{align}
The coefficients $A_n, B_n$ and $C_n$ are given in Appendix \ref{App:order1_solution}.
This gives us the rotational-flow-induced deformation $h^{r(1,0)}$,
\begin{align}
 \begin{split}
   h^{r(1,0)} = \left[C_1 K_1(\sqrt{\mbox{Bo}} r )   +\right. &  K_1(\sqrt{\mbox{Bo}} r ) \int_1^r r F^{r}(r) I_1(\sqrt{\mbox{Bo}} r ) \mbox{ d}r \\
    + &  \left. I_1(\sqrt{\mbox{Bo}} r ) \int_r^\infty r  F^{r}(r) K_1(\sqrt{\mbox{Bo}} r ) \mbox{ d}r\right]\cos\varphi,
\end{split}\label{eqn:app:rot_deformation}
 \end{align}
 with 
 \begin{align}
     C_1 = \frac{I_2(\sqrt{\text{Bo}}) \int_1^\infty r F^{r}(r) K_1(\sqrt{\text{Bo}}r)\mbox{ d}r}{K_2(\sqrt{\text{Bo}})}.
 \end{align}

\begin{figure}
    \centering
    \includegraphics[scale=0.6]{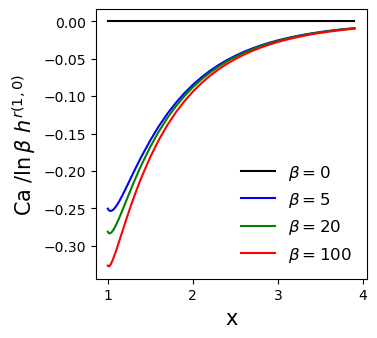}
    \caption{The scaled rotational-flow-induced deformation Ca$h^{r(1,0)}/\ln\beta$ is plotted along the positive $x$-axis with $\text{Ca} = 1, \text{Bo} = 1$, and varying values of slip coefficient $\beta$. }
    \label{fig:app:rot_deformation_against_r}
\end{figure}

Figure \ref{fig:app:rot_deformation_against_r} shows the interfacial profile of  Ca$h^{r(1,0)}$ scaled by $\ln\beta$ along the positive $x$-axis with Ca = 1, Bo = 1, and varying values of slip coefficient $\beta.$ The interface appears undeformed by the rotational flow when the particle surface has a perfect slip ($\beta = 0$). As $\beta$ increases, we observe  logarithmic growth in the deformation amplitudes. This unbounded growth is expected. The zeroth order rotational flow field $\mathbf{u}$ obtained by Ref. \cite{oneill1986} is valid under the assumption of a finite slip, i.e., $\beta \ll \infty.$ By letting $\beta \rightarrow \infty,$ the no-slip condition is recovered. This gives rise to a stress singularity at the TCL, i.e., the normal-normal stress $\sigma_{zz}^{r(0,0)}$ at $(r,z) = (1,0)$ diverges logarithmically as $\beta \rightarrow \infty$ \cite{oneill1986}.

\begin{figure}
    \centering
    \includegraphics[scale=0.6]{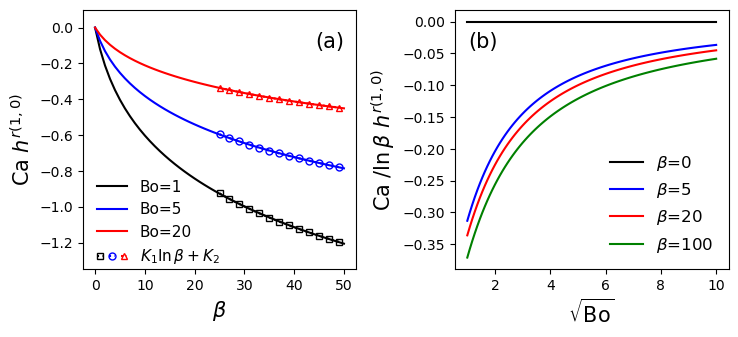}
    \caption{The interfacial height of Ca$h^{r(1,0)}$ at $(r,\varphi) = (1,0)$ plotted as functions of slip coefficient $\beta$ and  $\sqrt{Bo}$ for Ca = 1: (a) comparison of the interfacial heights Ca$h^{r(1,0)}$ and its large $\beta$-asymptotic approximation with Bo = 1, 2, 20, where $K_1 = -0.40$ and $K_2 = 0.37$ for $\text{Bo} = 1$, $K_1 = -0.27$ and $K_2 = 0.28$ for $\text{Bo} = 5$, and $K_1 = -0.16$ and $K_2 = 0.19$ for $\text{Bo} = 20$, (b) the scaled interfacial height Ca$h^{r(1,0)}/\ln\beta$ with $\beta = 0, 5, 20$, and 200. }
    \label{fig:app:rot_deformation_against_beta_Bo}
\end{figure}

In Fig. \ref{fig:app:rot_deformation_against_beta_Bo}, we plot the interface height Ca$h^{r(1,0)}$ at $(r,\varphi) = (1,0)$ as a function of the slip coefficient $\beta$ with Ca and varying values of Bond number Bo. Figure \ref{fig:app:rot_deformation_against_beta_Bo}(a) shows the logarithmic growth of the interface height Ca$h^{r(1,0)}$ as $\beta \rightarrow \infty,$ i.e., 
\begin{align}
     h^{r(1,0)}(1,\varphi) \sim K_1 \ln\beta + K_2,\quad \beta\gg1.\label{eqn:app:asym_approx_interface_height}
\end{align}
For large $\beta$, the coefficients $K_1$ and $K_2$ can be determined by curve fitting for a fixed value of Bo. In Fig.  \ref{fig:app:rot_deformation_against_beta_Bo}(a), we compare the exact solution of the interface height $h^{r(1,0)}$ at $r  =1$ and $\varphi = 0$ with the asymptotic approximation \ref{eqn:app:asym_approx_interface_height}, where $K_1 = -0.40$ and $K_2 = 0.37$ for $\text{Bo} = 1$, $K_1 = -0.27$ and $K_2 = 0.28$ for $\text{Bo} = 5$, and $K_1 = -0.16$ and $K_2 = 0.19$ for $\text{Bo} = 20$. Equation \eqref{eqn:app:asym_approx_interface_height} implies that for the correction deformation Ca$h^{r(1,0)}$  to be valid, we require $\text{Ca} \ll 1/\ln\beta. $
In Fig. \ref{fig:app:rot_deformation_against_beta_Bo}(b), we plot the scaled interface height Ca$h^{r(1,0)}/\ln\beta$ as a function of $\sqrt{\text{Bo}}.$ Here, the dependence of Ca$h^{r(1,0)}/\ln\beta$ on the Bond number is observed to be qualitatively the same as that of the interface height of the translational-flow-induced deformation. As $\sqrt{\text{Bo}}$ increases, the interface shape becomes flatter due to the increasing liquid density, and as shown in Fig. \ref{fig:app:rot_deformation_against_beta_Bo}(b), the interface height decreases.

\section{Far-field approximations of the interfacial deformations}\label{App:far-field_approx_of_flow_induced_defs}
The asymptotic approximation of the deformed interface around a translating or rotating particles is given by 
\begin{align}
    h\approx \text{Ca} h^{t,r(1,0)} + \delta h^{(0,1)},
\end{align}
Since the static deformation $\delta h^{(0,1)}$ far away from the particle decays exponentially, i.e., 
\begin{align}
    \delta h^{(0,1)} = - \delta \tilde{\theta}_s C_0 K_0(\sqrt{\text{Bo}} r) \sim -\delta \tilde{\theta}_s C_0\sqrt{\frac{\pi}{2}} \frac{e^{-\sqrt{\text{Bo}} r }}{\text{Bo}^{1/4} r^{1/2}},
    \label{eq:G2}
\end{align}
the interfacial deformation, $\text{Ca} h^{r,t(0,1)}$, induced by the particle's translational or rotational motion dominates in the far field. In the following sections, we calculate the far-field approximations of the flow-induced deformations $\text{Ca} h^{r,t(0,1)}$.

\subsection{Deformation induced by particle translation}
We first consider translational flow-induced deformation $\text{Ca}h^{t(1,0)}$,  which is given in Eq. \eqref{eqn:flow_induced_deformations_solutions} and can be expanded as
\begin{align}
\begin{split}
\text{Ca}h^{t(1,0)} = \text{Ca}\left[C_1 K_1(\sqrt{\mbox{Bo}} r )   +\right. & \frac{3\beta}{2(3+\beta)} K_1(\sqrt{\mbox{Bo}} r ) \int_1^r \frac{1}{r^3} I_1(\sqrt{\mbox{Bo}} r ) \mbox{ d}r \\
    + &  \left. \frac{3\beta}{2(3+\beta)} I_1(\sqrt{\mbox{Bo}} r ) \int_r^\infty \frac{1}{r^3} K_1(\sqrt{\mbox{Bo}} r ) \mbox{ d}r\right]\cos\varphi.\label{eqn:app:trans_flow_induced_deformation}
    \end{split}
\end{align}
As $r\rightarrow \infty,$ the first term in the brackets in  Eq. \eqref{eqn:app:trans_flow_induced_deformation} decays exponentially, i.e., 
\begin{align}
    C_1 K_1(q r) \sim C_1 \sqrt{\frac{\pi}{2} }  \frac{ e^{-\sqrt{\text{Bo}} r }}{\text{Bo}^{1/4} r^{1/2}},
\end{align}
and the asymptotic approximation of the second and third terms in the brackets in Eq. \eqref{eqn:app:trans_flow_induced_deformation} is obtained to be 
\begin{align}
     K_1(\sqrt{\mbox{Bo}} r ) \int_1^r \frac{1}{r^3} I_1(\sqrt{\mbox{Bo}} r ) \mbox{ d}r 
    +  \frac{3\beta}{2(3+\beta)} I_1(\sqrt{\mbox{Bo}} r ) \int_r^\infty \frac{1}{r^3} K_1(\sqrt{\mbox{Bo}} r ) \mbox{ d}r \sim \frac{1}{\text{Bo} r^4}. 
\end{align}
This gives us the leading-order far-field approximation of $h^{t(1,0)}$,
\begin{align}
\text{Ca}h^{t(1,0)} \sim \frac{\text{Ca} 3\beta}{2(3+\beta)} \frac{1}{\text{Bo} r^4} \cos\varphi. 
\label{eq:G6}
\end{align}
Note that this approximation is valid for $\sqrt{\text{Bo}} r \gg 1.$
    
\subsection{Deformation induced by particle rotation} 
The interfacial deformation induced by the particle's rotational motion can be obtained by paralleling the calculation of the translational-flow-induced-deformation in Sec. \ref{sec:interfacial_deformations}. 
We compute the leading-order normal-normal stress $\sigma_{zz}^{r(0,0)}$ at the interface $\Sigma_I^{(0,0)}$ from the rotational flow field $\mathbf{u}^{r(0,0)}$ obtained by O'Neill \textit{et al.} \cite{oneill1986}, i.e., 
\begin{align}
     \sigma_{zz}^{r(0,0)}  =  F^r(r)\cos\varphi,
\end{align}
with 
    \begin{align}
    F^r(r) =  \sum_{n=0}^\infty -\frac{4\sqrt{\pi}  (A_n (2n+1)^2 (2n+3) + 2n(C_n +2 B_n (2n^2 + n-1) )  r^2 ) }{\Gamma(-1/2-n)\Gamma(n+1) r^{2(n+2)}}. 
\end{align}
Note that at the TCL, the normal-normal stress $\sigma_{zz}^{r(0,0)} \sim \log(r-1)$ \cite{oneill1986}.
Then, the rotational-flow-induced deformation $h^{r(1,0)}$ is given by 
\begin{align}
 \begin{split}
   \text{Ca}h^{r(1,0)} = \text{Ca}\left[C_1 K_1(\sqrt{\mbox{Bo}} r )   +\right. &  K_1(\sqrt{\mbox{Bo}} r ) \int_1^r r F^{r}(r) I_1(\sqrt{\mbox{Bo}} r ) \mbox{ d}r \\
    + &  \left. I_1(\sqrt{\mbox{Bo}} r ) \int_r^\infty r  F^{r}(r) K_1(\sqrt{\mbox{Bo}} r ) \mbox{ d}r\right]\cos\varphi,
\end{split}
 \end{align}
 with 
 \begin{align}
     C_1 = \frac{I_2(\sqrt{\text{Bo}}) \int_1^\infty r F^{r}(r) K_1(\sqrt{\text{Bo}}r)\mbox{ d}r}{K_2(\sqrt{\text{Bo}})}.
 \end{align}
Second, we consider the rotational-flow-induced deformation $h^{r(1,0)}$ given in Eq. \eqref{eqn:app:rot_deformation}. The function $F^r(r)$ in Eq. \eqref{eqn:app:rot_deformation} is given as an infinite series in $r$, which can be expended as 
\begin{align}
 F^r(r)   
         = & \frac{6 A_0 }{r^4} + \left( -\frac{24 B_1 + 6 C_1}{r^4} - \frac{135 A_1}{r^6}\right) + \cdots.
\end{align}
Truncating after the asymptotically dominant term, we obtain the leading-order approximation of $F^r(r)$ for large $r,$ 
\begin{align}
    & F^r(r) \sim \frac{6A_0 - 24 B_1 - 6 C_1}{r^4}  = -\frac{3\beta (7\beta^2 + 83 \beta + 210)}{8(\beta+3)(\beta+4)(\beta+7) r^4}= \frac{\mathcal{D(\beta)}}{r^4}.
\end{align}
This yields the far-field approximation of $h^{r(1,0)}$,
\begin{align}
    \text{Ca}h^{r(1,0)} \sim \frac{\text{Ca} \mathcal{D}(\beta)}{\text{Bo} r^4} \cos\varphi,\quad r\gg 1. 
\end{align}

\section{Capillary interaction forces for $\text{Bo} = 0$}
As discussed in Appendix \ref{App:asymptitics_bond_number} and Ref. \cite{zhou2022}, the static deformation $\delta h^{(0,1)}$ for Bo = 0 fails to satisfy the contact angle condition \eqref{eqn:far_field_condition_static_def} for contact angle $\tilde{\theta}_s \neq 90^\circ.$ Here, for the purpose of analysis, we assume $\theta_s = 90^\circ (\tilde{\theta}_s = 0)$, which implies $\delta h^{(1,0)} \equiv 0$, and only consider the capillary effects due to the flow-induced deformations $\text{Ca} h^{t,r(1,0)}$
For $\mbox{Bo} = 0,$ the flow-induced deformations $h^{t,r(1,0)}$ and their leading-order approximations for $r \gg 1$ are given by 
\begin{align}
 &   h^{t(1,0)} = \frac{3}{2(\beta+3)} \left(-\frac{1}{2r} + \frac{1}{3r^2}\right)\cos\varphi \sim -\frac{3}{4(\beta+3) } \frac{1}{r}\cos\varphi,\\
   &   h^{r(1,0)} = \left(-\sum_{n=0}^\infty \frac{\tilde{\mathcal{D}}_n}{2(2n+1)} \frac{1}{r} + \sum_{n=0}^\infty \frac{\tilde{\mathcal{D}}_n}{4n^2+8n+3} \frac{1}{r^{2n+2}} \right)\cos\varphi \sim -\frac{\tilde{\mathcal{D}}_0}{2}\frac{1}{r} \cos\varphi\\
\end{align}
with
\begin{align}
\tilde{\mathcal{D}}_n = 4\sqrt{\pi}\left(\frac{A_n(2n+1)^2 (2n+3)}{\gamma(-1/2-n)\gamma(n+1)} + \frac{2(n+1)C_{n+1}+2B_{n+1}(2(n+1)^2+n)}{\gamma(-3/2-n)\gamma(n+2)}\right).
\end{align}
The LSA of the two-particle deformations can be calculated from the asymptotically dominant terms in the single-particle deformations $\text{Ca}h^{t,r(1,0)}$. Then, paralleling the calculations in Sec. \ref{subsec:trans:capillary_interactions},  we obtain the nondimensionalized leading-order lateral interaction forces exerted on two translating and rotating particles,
\begin{align}
    \mathbf{F}_{int}^t \sim & \frac{9\text{Ca} \pi \beta^2 \cos\Phi\sin^2\Phi }{(3+\beta)^2 L^3} \hat{\mathbf{e}}_x + \frac{9\text{Ca}\pi\beta^2 (\sin\Phi-\sin 3\Phi)}{4(3+\beta^2)L^3} \hat{\mathbf{e}}_y,\\
    \mathbf{F}_{int}^r \sim & \frac{4\text{Ca} \pi \tilde{\mathcal{D}}_0^2 \cos\Phi\sin^2\Phi }{L^3} \hat{\mathbf{e}}_x + \frac{\text{Ca}\pi \tilde{\mathcal{D}}_0^2 (\sin\Phi-\sin 3\Phi)}{L^3} \hat{\mathbf{e}}_y. 
\end{align}

\bibliography{references}

\end{document}